\documentclass[sn-standardnature]{sn-jnl}
\usepackage{graphicx,subfigure}

\jyear{2022}%

\theoremstyle{thmstyleone}%
\theoremstyle{thmstyletwo}%

\theoremstyle{thmstylethree}%

\begin{document}

\title[Information scrambling and redistribution of quantum 
correlations]{Information scrambling and redistribution of quantum correlations 
through dynamical evolution in spin chains}


\author*[1]{\fnm{Saikat} \sur{Sur}}\email{saikat.sur@weizmann.ac.il}

\author[2]{\fnm{V.} \sur{Subrahmanyam}}\email{vmanisp@uohyd.ac.in}


\affil*[1]{\orgdiv{Department of Chemical and Biological Physics}, 
\orgname{Weizmann Institute of Science}, \city{Rehovot}, \postcode{7610001}, 
\country{Israel}}

\affil[2]{\orgdiv{School of Physics}, \orgname{University of Hyderabad}, 
\city{Hyderabad}, \postcode{500046}, \country{India}}


\abstract{We investigate the propagation of local bipartite quantum correlations, along with
the tripartite mutual information to characterize the information scrambling through
dynamical evolution of spin chains. Starting from an initial state with the first pair
of spins in a Bell state, we study how quantum correlations spread to other parts
of the system, using different representative spin Hamiltonians, viz. the Heisenberg
Model, a spin-conserving model, the transverse-field XY model, a non-conserving
but integrable model, and the kicked Harper model, a spin conserving but noninte-
grable model. We show that the local correlations spread consistently in the case of
spin-conserving dynamics in both integrable and nonintegrable cases, with a strictly
nonnegative tripartite mutual information. In contrast, in the case of nonconserving
dynamics, tripartite mutual information is negative and local pair correlations do not
propagate.}





\maketitle

\section{Introduction}\label{sec1}
Quantum entanglement and information in the ground state and  the dynamics of 
quantum spin chains have been extensively studied over the last few years, as 
spin chains are perceived to be possible channels  for quantum communication and 
information processing. Traditionally, quantum spin chains have been 
investigated from the view point of quantum state and entanglement 
transfer~\cite{bose,subra1,christandl,wang,subrahmanyam_2006}, studying  
low-dimensional condensed matter physics systems exhibiting  quantum phase 
transitions and a variety of spin orderings~\cite{sachdev,jordan,lieb, 
takahashi,bethe}. These systems have also been studied in the context of the 
dynamics of quantum many-body systems, for magnon bound and scattering 
states~\cite{ganahl,fukuhara,vlijm},  spin current dynamics~\cite{steingeweg} 
and relativistic density-wave dynamics~\cite{foster}. The unitary evolution of 
quantum correlations, using model Hamiltonians, has been investigated in various 
fields, e.g.,  quantum quench dynamics~\cite{cazalilla,mukherjee,nag}, the 
light-cone entanglement spreading~\cite{manmana,sodano,chiara}, dynamics of 
disordered systems~\cite{safavi, roeck}. 

In the last few years, scrambling of quantum information has been investigated 
in various contexts, for example, in information paradox in black  
holes~\cite{ioyada_17,ioyada_18}, in many-body disordered systems \cite{ref1}, 
in understanding the dynamics of  ergodic and integrable systems~\cite{campisi, 
landsman, seshadri}.    Now, scrambling of information is quantified by an 
observable independent quantity called tripartite mutual information 
(TMI)~\cite{scram1, scram2}, a combination of bipartite and tripartite quantum 
correlation functions. 
  It is interesting to investigate the connection between TMI,  that is the 
information scrambling,  with the bipartite quantum correlations in a dynamical 
many-body system. Since the dynamical evolution of the spin chain is through a 
Hamiltonian evolution, the symmetry property of the spin Hamiltonian and the 
integrability of the underlying dynamics are expected to play a significant role 
on the scrambling nature of the dynamics. In this paper, we will address these 
by studying the dynamics of multi-qubit states initialized with a locally 
encoded information or an entangled pair.  We will be investigating the dynamics 
of TMI and connect it  with the dynamics of two popular bipartite quantum 
correlation functions, the concurrence measure of quantum entanglement and the 
quantum mutual information.
  
  We use different spin models, spin conserving or spin non-conserving, 
integrable or nonintegrable, for the spin chain dynamics.  We study the 
Heisenberg model, that exhibits a spin-conserving dynamics, the XY model in a 
transverse field with a spin non-conserving dynamics and  the  kicked Harper 
model with a non-integrable but spin-conserving dynamics. We investigate how the 
entanglement is generated and distributed over various pairs of qubits,  by 
studying the time dependence of concurrence measure of pairwise entanglement and 
pairwise quantum mutual information. The time evolution of tripartite mutual 
information can be used for each of these cases  to categorize the dynamics of 
spin chains on the basis of their scrambling behaviour, and the relationship 
between propagation of bipartite correlations and scrambling of information. 
This paper is organised as follows. We discuss measures of bipartite 
correlations and scrambling of quantum information, that involves one-party, 
two-party and three-party correlation functions in section~\ref{section_2}. We 
study the time evolution of one-magnon and two-magnon initial states and their 
correlation dynamics, for the Heisenberg model, the kicked Harper model and the 
transverse-field XY model and the corresponding dynamics of TMI in 
section~\ref{section_3}.

\begin{figure}[h]
\begin{center}
        \subfigure[]{%
            \includegraphics[width=0.33\textwidth]{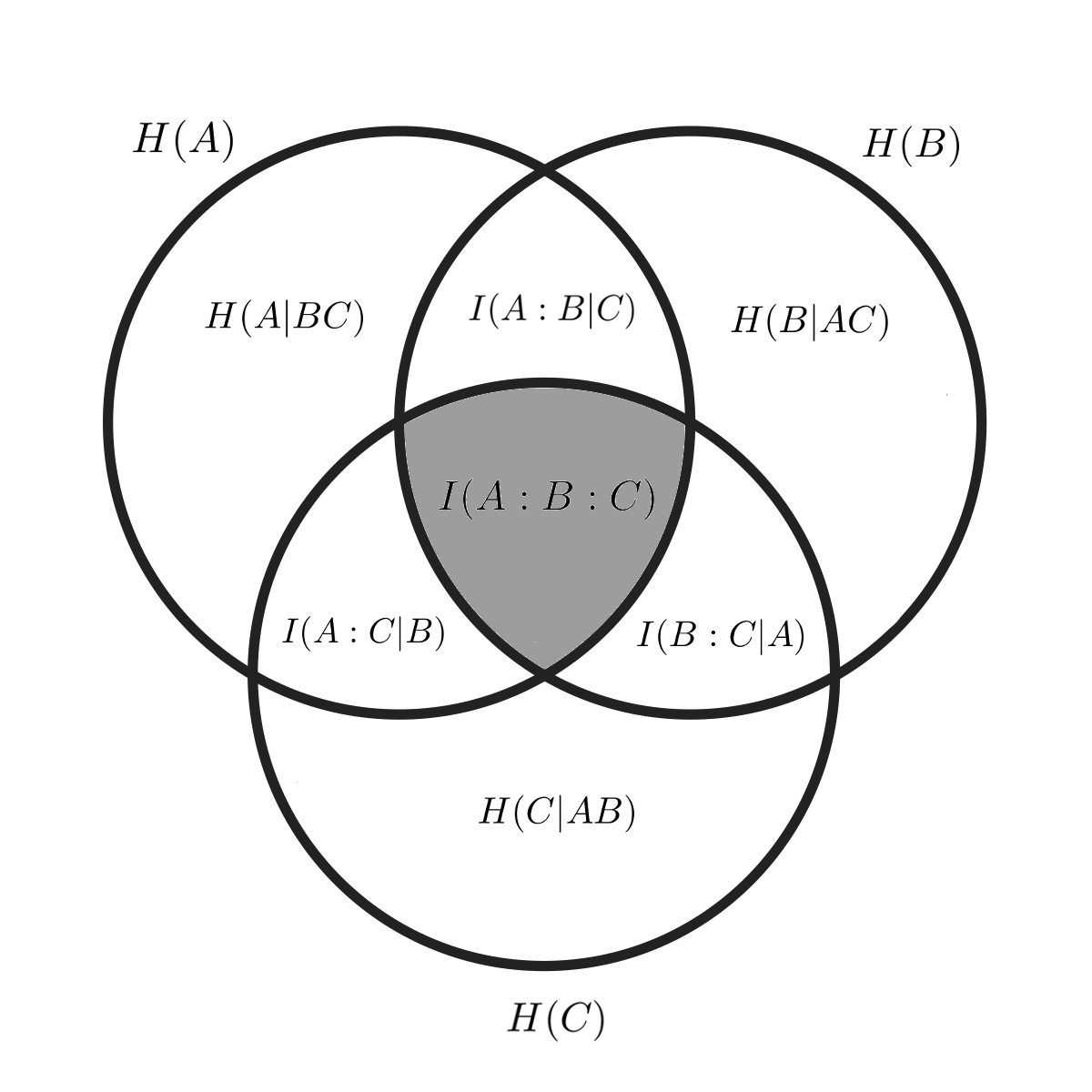}
        }%
        \subfigure[]{%
           \includegraphics[width=0.35\textwidth]{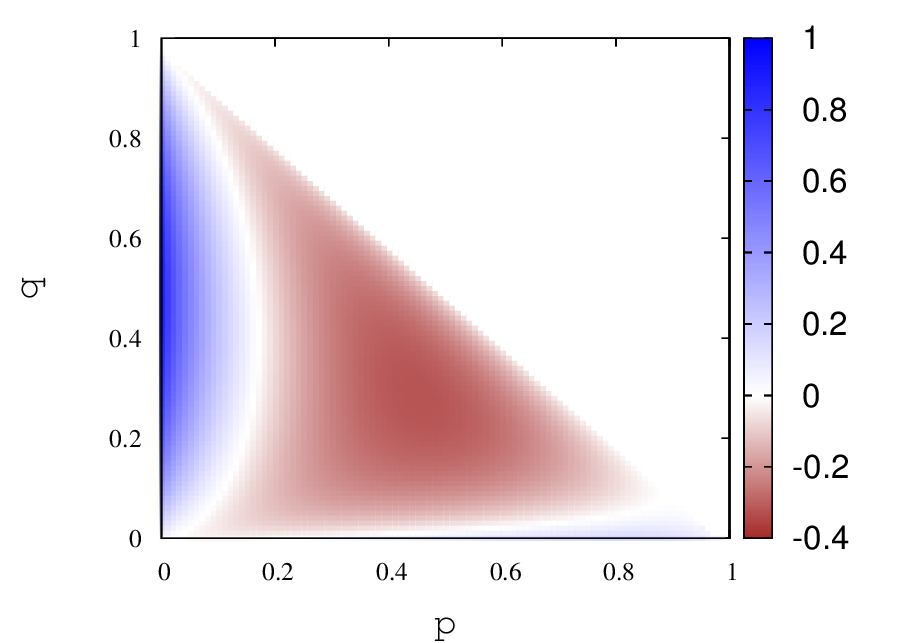}
        }\\%
\end{center}
\caption{{(a) The information diagram for three classical variables $A$, $B$ and 
$C$ with the shaded region denoting the tripartite mutual information 
$I(A:B:C)$. (b) The density plot of the tripartite mutual information  
$I(1:2:3)$  between three spins for the state given in Eq.~\ref{mixed} as a 
function of $p$ and $q$.  }}
\label{fig:tmi}
\end{figure}

\section{Quantum correlations and scrambling of quantum information}
\label{section_2}

We briefly review a few popular measures of quantum correlations and information 
that have been extensively studied for spin systems over the last two decades. 
The pairwise entanglement and the mutual information will depend on two-point 
correlation functions. {The two-qubit reduced density matrix  $\rho_{AB}$ of two 
qubits $A$ and $B$ is computed by tracing out all other qubits from the full 
system. In general, for a many-body system, the two-qubit state $\rho_{AB}$ will 
be a mixed state.} The dynamics, for all the models that we will be considering 
below, conserves the parity, which implies that the  elements of the reduced 
density matrix between two states with odd and even number of up (down) spins 
are zero. \\

Let us consider $\sigma^z$ - diagonal basis states, $\vert 00\rangle$,  $\vert 
01\rangle$, $\vert 10\rangle$, $\vert 11\rangle$, for two qubits. The two-qubit 
reduced density matrix   $\rho_{AB}$ has the  $X$-state form, given as

\begin{eqnarray}
\rho_{AB}=  \begin{pmatrix}  u & 0 & 0 & z \\ 0 & {w_1} & x & 0 \\ 0 & {x^*} & 
{w_2} & 0\\ {z^*} & 0 & 0 & v  \end{pmatrix}
  \label{two_party_rdm}
  \end{eqnarray}
The above matrix elements are related to two-point diagonal and off-diagonal 
correlation functions~\cite{subrah-heisen}. For spin-conserving models, like the 
Heisenberg dynamics, the off-diagonal matrix element $z$ is zero if the initial 
state has a definite number of down (up) spins, as states with different down 
spins do not mix. For spin non-conserving transverse-field XY model, the matrix 
element $z$ can be nonzero~\cite{aritra}. {The pairwise concurrence, between two 
marked qubits, quantifies their mutual entanglement, that takes the value of 
unity when they are maximally entangled and zero when they are separable. The 
concurrence \cite{wootters} measure is defined as,}
  \begin{equation}
C =  \mbox {max} \{0,\sqrt{\lambda_1} -\sqrt{\lambda_2} -\sqrt{\lambda_3} 
-\sqrt{\lambda_4} \}.
\end{equation}
Here, $\lambda_i$ are eigenvalues of the matrix $\rho_{AB} \tilde\rho_{AB}$ in 
decreasing order, where $\tilde\rho_{AB}$  is the time reversed state, given by 
$\tilde\rho_{AB} = (\sigma_A^y \otimes \sigma_B^y) \rho_{AB}^* (\sigma_A^y 
\otimes \sigma_B^y)$. The concurrence between two qubits, in terms of  the two 
qubit RDM elements is given by the following form~\cite{subra1},
\begin{equation}
C = 2~\mbox {max} \{0,\vert x \vert -\sqrt{uv},\vert z \vert -\sqrt{{w_1}{w_2}} 
\}.
\end{equation}

The quantum mutual information for two parties $A$ and $B$ in the quantum state 
$\rho_{AB}$, is defined in terms of the von Neumann entropies $S(\rho_A)$ and 
$S(\rho_B)$ and $S(\rho_{AB})$ (where $S(\rho)=-Tr \rho \log_2 \rho$), given by
\begin{equation}
I(A:B) = S(\rho_A) + S(\rho_B) -S(\rho_{AB}).
\label{mutual_inf}
\end{equation}
Unlike the concurrence measure, the mutual information is related to the 
eigenvalues of the single and  the two-qubit reduced density matrices.
By the concavity property of von Neumann entropy the mutual information is a 
non-negative quantity and bounded between $0$ and $2$. Clearly, the mutual 
information is zero for pure bipartite factorizable states, as both the 
composite-system states and the states of its sub-parts are pure and there is no 
loss of information in accessing the composite system locally.
where we can have entanglement within pure states.

Before discussing tripartite mutual information for quantum systems in detail, 
let us briefly outline the concept of tripartite mutual information in classical 
information theory. If $A$, $B$ and $C$ are three random variables, the TMI 
$I(A:B:C)$ is defined as~\cite{thomas_cover},
\begin{equation}
I(A:B:C) = I(A:B) +I(A:C) -I(A:BC).
\end{equation}
TMI is not necessarily non-negative unlike the case of mutual 
information with two variables. The value of $I(A:B:C)$ between three classical 
variables $A$, $B$ and $C$ is $-1$ if  $B$ and $C$ are independent random 
variables and $A = B \oplus C$. In this case, the information shared between the 
random variables $A$ and $BC$ jointly is larger than the informations shared 
between the random variables $A$ and $B$; $A$ and $C$ jointly. The value of 
$I(A:B:C)$ between them is $1$ if they all are identical, i.e., $A = B = C$. On 
the other hand, $I(A:B:C) = 0$ if all the variables $A$, $B$ and $C$ are 
independent and random. 

In the context of quantum dynamics, the locally-encoded quantum information 
spreads out over the entire system through a unitary time evolution. Such 
delocalization of quantum information is referred to as scrambling \cite{ref1, 
scram1, scram2, ioyada_17,ioyada_18}. This implies that local disturbances in 
the initial states cannot be detected by local measurements on the output states 
if scrambling of information occurs. Let us consider a system consisting of 
three subparts $A$, $B$ and $C$. In this case, a local measurement on a subpart 
$A$ may not reveal much information about a local disturbance at a different 
subpart $B$ if scrambling occurs.  Therefore, the mutual information $I(A : B)$  
would be small. From a similar reasoning, the pairwise mutual information $I(A : 
C)$ also would be small. The mutual information $I(A : BC)$ quantifies the total 
amount of information one can learn about $A$ by measuring the part $BC$ 
jointly. Since one is interested in the amount of information concerning $A$, 
which is hidden non-locally over $B$ and $C$, a  measure of scrambling would be 
given by the quantity known as tripartite mutual information (TMI) 
\cite{scram1,scram2}. This is a measure to quantify  by how much the information 
shared by $A$ with $B$ and $C$ together is different from the sum of the 
informations shared pairwise by $A$ with $B$, and  by $A$ with $C$. Using the 
definition of the mutual information given in Eq.~\ref{mutual_inf}, one can 
rewrite the expression fot TMI in terms of the von Neumann entropies of one, two 
 and three-party reduced density matrices, symmetric in the labels  $A$, $B$, and 
$C$ as
\begin{eqnarray}
I(A:B:C)=S(\rho_A)+S(\rho_B)+ S(\rho_C)- S(\rho_{ AB}) - S(\rho_{ BC})- S(\rho_{ 
CA})+ S(\rho_{ ABC}). \nonumber\\
\end{eqnarray}

If $\rho_{ABC}$ represents a pure state, for any partition 
between the subparts $A$ and $BC$, the two parts will have the same von Neumann 
entropy, i.e., $S(\rho_A) = S(\rho_{BC})$, and similarly for other 
bi-partitions. Thus, the tripartite mutual information is identically zero.  But 
it becomes a non-trivial measure for a tripartite mixed state or a many-qubit 
pure state with at least four qubits. A positive TMI implies that $A$ with $B$ 
and $A$ with $C$ share some redundant information as compared to $A$ with $BC$. 
In this case, quantum correlations and information between two parties $A - B$ 
and $A - C$ propagate to other qubits in a many-qubit system through further 
time evolution as a form of bipartite correlations. On the other hand,  a  
negative TMI implies that the sum of information shared between $A$ and $B$ ; 
$A$ and $C$ is smaller than that between $A$ and $BC$ together, signifying that 
a fraction of quantum correlations are delocalized in the form of correlations 
among three or more parties as opposed to local two-party correlations.  As a 
consequence, this delocalized information do not propagate as bipartite 
correlations through further time evolution with a finite speed through the 
system. This is a signature for the loss of information, and hence scrambling of 
information. Evidently, the more the negative TMI is, the more is the 
delocalization of bipartite correlations. 

To illustrate the scrambling of information and TMI,  let us consider a few 
representative states. For a four-qubit pure state of the form,  $\vert 
\psi\rangle =  \vert \phi\rangle_1 \otimes \vert \phi\rangle_{234}$, we have  
$I(1:2:3) = 0$, where $1$ and $234$ are not entangled. For the four-qubit 
GHZ-like state, $\vert \psi\rangle = (\vert 0000 \rangle + \vert 
1111\rangle)/\sqrt{2}$ has  $I(1:2:3) = 1$ for any three qubits. For the 
four-qubit W-state $\vert \psi\rangle = (\vert 1000 \rangle + \vert 0100\rangle 
+ \vert 0010 \rangle + \vert 0001\rangle)/2$ has  $I(1:2:3) \sim 0.244$ for any 
three qubits.  Now, to illustrate that TMI can be negative, let us consider a 
simple three-qubit mixed state given as,
\begin{eqnarray}
 \rho = p \frac{1}{3}\vert 100 + 010 +001\rangle \langle 100 + 010 +001 \vert + 
q \vert 111\rangle \langle 111 \vert + (1-p-q)\vert 000\rangle \langle 000 
\vert,\nonumber\\
 \label{mixed}
\end{eqnarray}
where, $p, q \ge 0$ and $p+q \le 1$.  For this state the tripartite mutual 
information is straightforward to compute, and  shown as a density plot in 
Fig.~\ref{fig:tmi}(b) as a function of $p$ and $q$. For  the cases $p = 0$, $q = 
1$ and $p =1$, $q = 0$ the state $\rho$ is a pure three qubit state and 
$I(1:2:3)$ becomes zero as evident from its definition. The sign of $I(1:2:3)$ 
depends on the values of $p$ and $q$ in the rest of the region. For instance, we 
can see that the tripartite mutual information for this state is negative for $p 
\gtrsim 0.2$.

\section{Dynamical evolution of local quantum correlations in various spin 
models}
\label{section_3}

We  now consider the dynamical evolution of an initial spin state using 
Hamiltonian dynamics. Through the time-evolution, a variety of spin correlations 
can be dynamically generated from a locally correlated initial state.  We  
explicitly study below, three model spin Hamiltonians: the anisotropic 
Heisenberg Hamiltonian, an integrable and spin-conserving model, the 
periodically kicked Harper model, a nonintegrable but spin-conserving model, 
and the transverse field  XY model, an integrable but spin non-conserving model. 
The propagation of a signal of quantum dynamical process and interference with 
the state transfer have been investigated for these 
models~\cite{saikat1,saikat2}, where it was seen that a finite speed of 
propagation cannot be defined for nonintegrable dynamics. We will see below 
that the consistent spreading of local quantum correlations does not depend on 
non-integrability, but depends significantly on whether the dynamics is spin 
conserving or not.

 \begin{figure*}[t]
     \begin{center}
        \subfigure[]{%
            \includegraphics[width=0.35\textwidth]{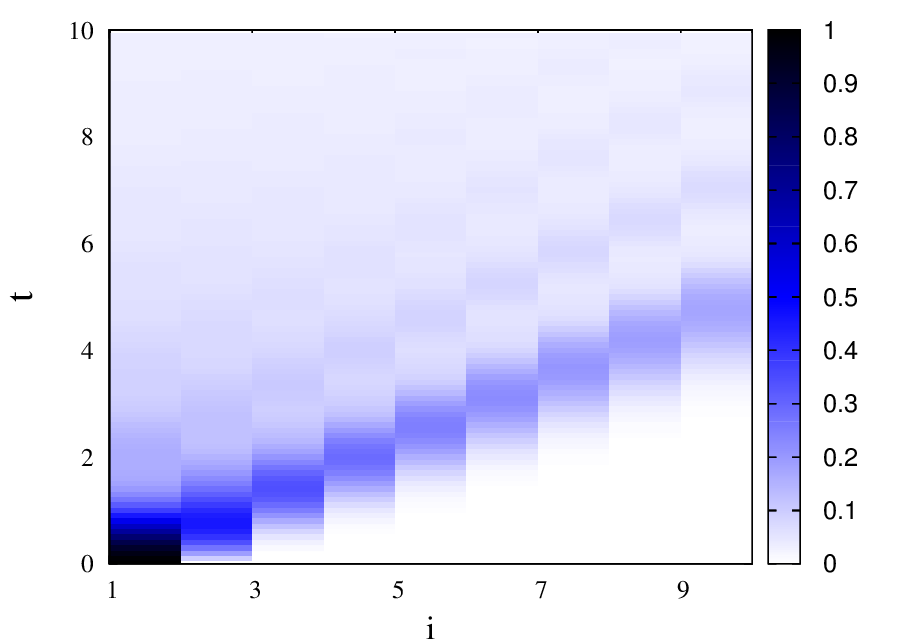}
        }%
        \subfigure[]{%
           \includegraphics[width=0.35\textwidth]{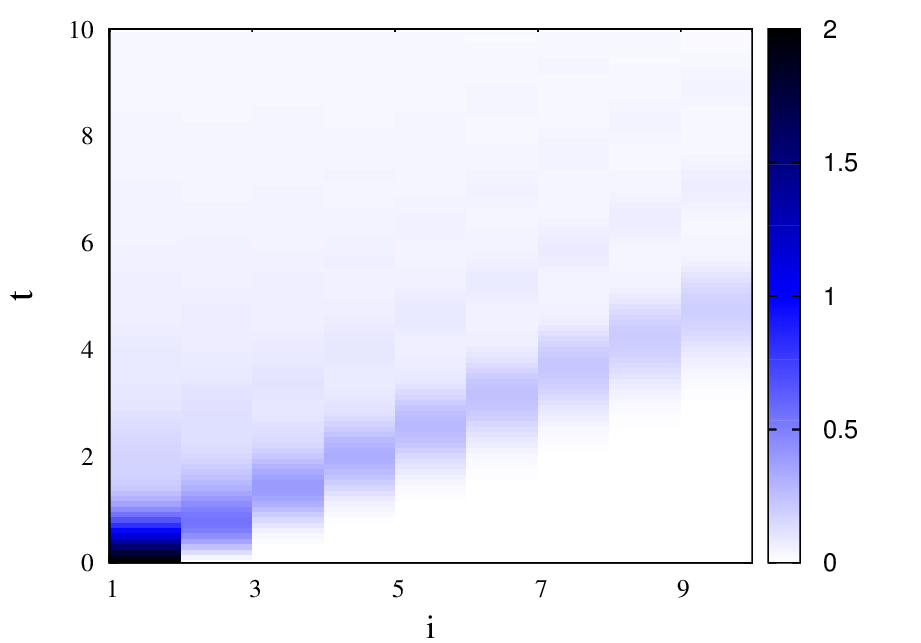}
        }\\%
        \subfigure[]{%
            \includegraphics[width=0.35\textwidth]{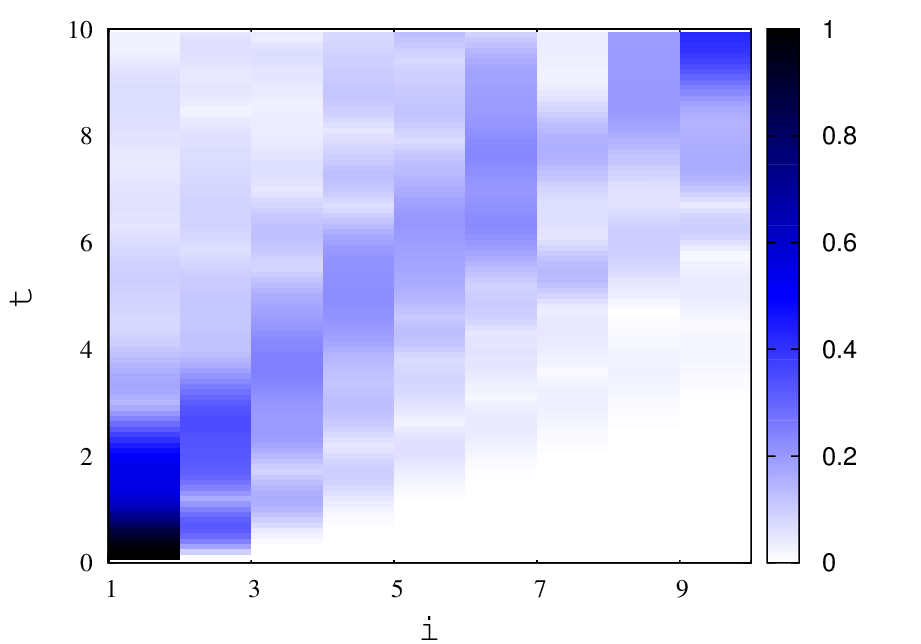}
        }%
        \subfigure[]{%
           \includegraphics[width=0.35\textwidth]{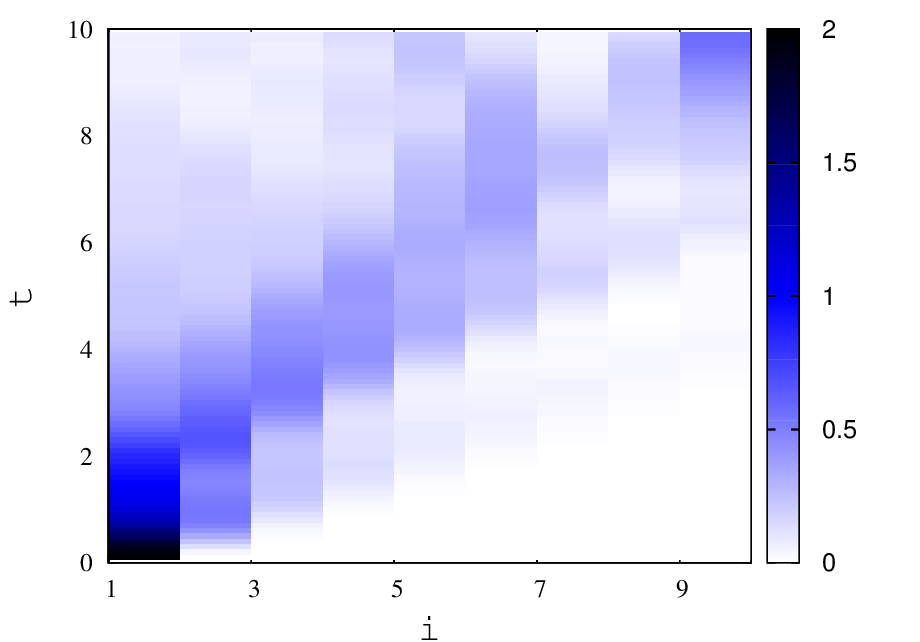}
        }%
    \end{center}
\caption{{The time dependence of the concurrence and the mutual information 
between the sites $i$ and $i+1$ as functions of time $t$ for the 
Heisenberg model: (a) the concurrence, (b) the mutual information for the initial 
state $ \vert10...0+010...0\rangle/\sqrt{2}$, (c) the concurrence,  (d) the 
mutual information for the initial state $ \vert00...0+110...0\rangle/\sqrt{2}$. 
The anisotropy constant is $\Delta = 1.0$. }     }%
   \label{fig:2}  
\end{figure*}

\begin{figure}[h]
\begin{center}
        \subfigure[]{%
            \includegraphics[width=0.35\textwidth]{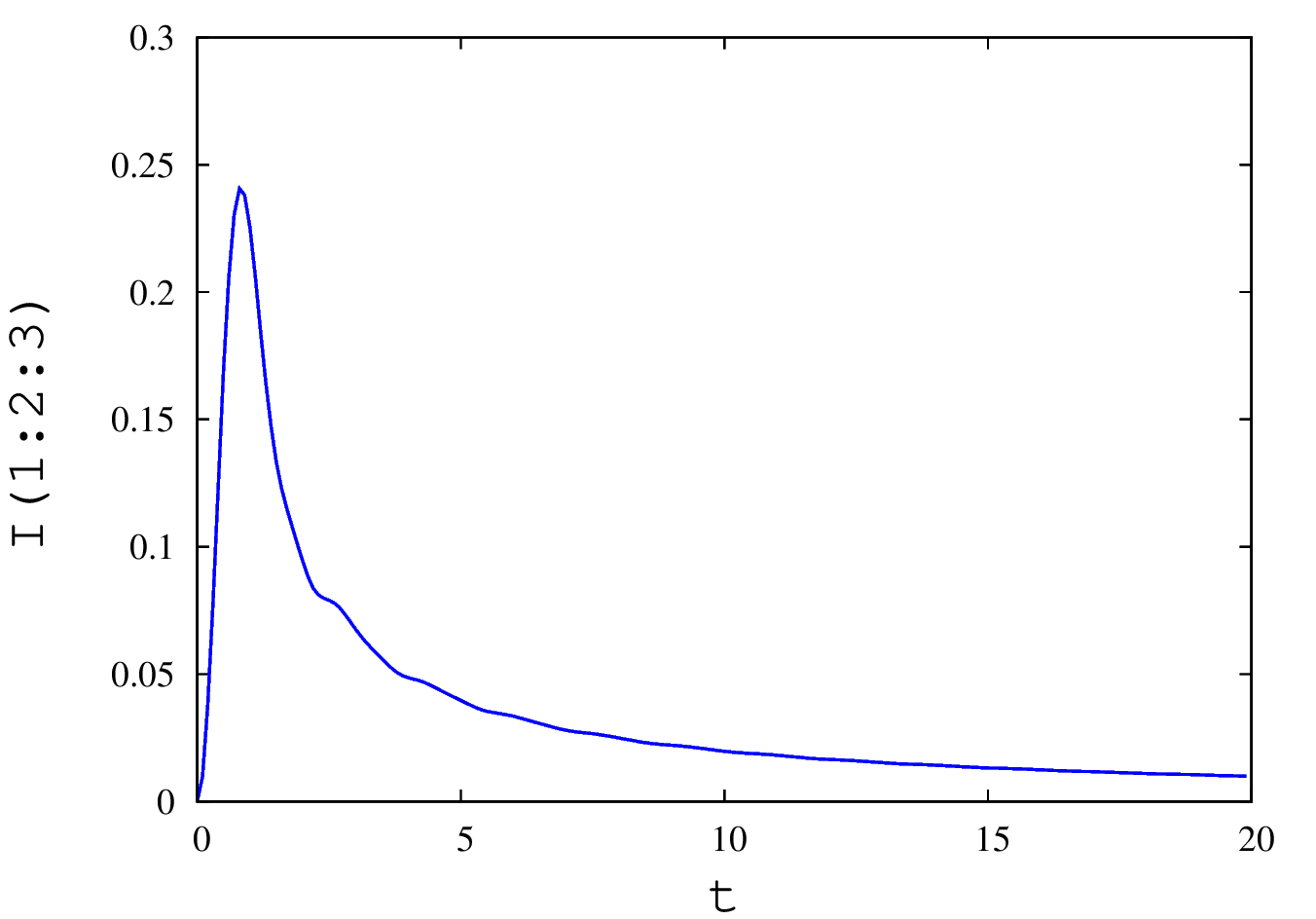}
        }%
        \subfigure[]{%
           \includegraphics[width=0.35\textwidth]{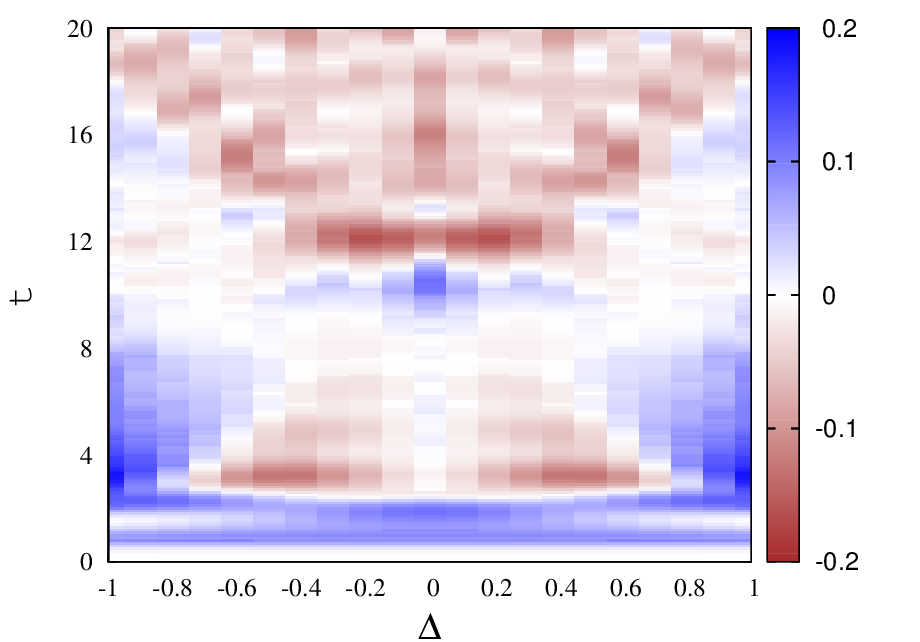}
        }\\%
\end{center}
\caption{{The time dependence of TMI $I(1:2:3)$ for anisotropic Heisenberg model 
for the initial states (a) $\vert01...0+100...0\rangle/\sqrt{2}$ and  (b) 
$\vert00...0+110...0\rangle/\sqrt{2}$   with anisotropy constant $\Delta$ with 
time $t$. }}
\label{fig:3}
\end{figure}

 \subsection{Anisotropic Heisenberg model}
 We  consider first an exactly-solvable and integrable non-trivial model of 
interacting quantum spins in a one dimensional array, interacting with 
nearest-neighbour Heisenberg exchange interaction. Using the Pauli matrices $ 
\vec \sigma_i $ to represent the different components of the spin operator at 
the $i$th site, the anisotropic Heisenberg Hamiltonian for a one-dimensional 
chain of $N$ spins is given by,
 \begin{equation}
H = 
-J\sum_{i}({\sigma}^x_{i}{\sigma}^x_{i+1}+{\sigma}^y_{i}{\sigma}^y_{i+1}+\Delta 
{\sigma}^z_{i}{\sigma}^z_{i+1}),
 \end{equation}
where $J$ is the exchange interaction strength for the nearest-neighbour spins, and $\Delta$ is the anisotropy parameter. 
The model exhibits ferromagnetic  behaviour in the ground state for $\Delta \ge 1$. In the region, $-1< \Delta <1$,  the model is gap-less and is in the 
spin-liquid phase with a power law decay of correlations.  In the region, $\Delta \le -1$, the ground state shows a N\'eel long range order. All the 
eigenstates of this model are known, and can be found using the Bethe Ansatz~\cite{bethe}. 

Let the basis states for the $i$th spin be $\vert 0\rangle$ (up-spin state) and  
$\vert 1\rangle$ (down-spin state), denoting the eigenstates of $\sigma_i^z$ 
with eigenvalues $+1$ and $-1$ respectively. The basis states for the many-qubit 
system can be chosen to be the direct product states of the basis states of each 
spin. The z-component of the total spin $\Sigma 
\sigma_i^z$ is a constant of motion,  which implies that the eigenstates will 
have a definite number of down spins and parity.  The many-qubit basis states 
with $l$ down spins can also 
be labeled by the locations ($x_1,x_2...x_l$), where the set is an ordered set 
with $x_1<x_2...<x_l$. An eigenstate with $l$ down spins, a $l$-magnon state,  
can be written as a superposition of the basis states as,
\begin{equation}
\vert\psi\rangle = \sum_{x_1,x_2...,x_l}\psi ({x_1,x_2...x_l}) \vert 
x_1,x_2...x_l\rangle,
\end{equation}
where the eigenfunction $\psi({x_1,x_2...,x_l})$ denotes the wave function 
amplitude for the  basis state $\vert x_1,x_2...,x_l\rangle$. The eigenfunction 
is given by the Bethe Ansatz~\cite{izyu}, labeled by the set of quasi-momenta 
$(p_1,p_2...p_l)$ of the down spins, which are determined by solving algebraic 
Bethe Ansatz equations, with periodic boundary conditions.  

There is only one zero-magnon state $\vert F\rangle =\vert 00...0\rangle$, which 
is just a ferromagnetic ground state with all the spins polarized along one 
direction. It is straightforward to see that it is an eigenstate of the above 
Hamiltonian with energy $\epsilon_0=-N J \Delta$ for periodic boundary 
conditions.  Starting from $\vert F\rangle$, one-magnon excitations can be 
created by turning any one of the spins, giving $N$ localized one-magnon states, 
which can be labeled by the location of the down spin. One-magnon eigenstates 
are labeled by the momentum  $p$ of the down spin, the eigenfunction is given 
by,
\begin{eqnarray}
\psi_p^x = & \sqrt{\frac{1}{N}} e^{i px}; p ={ 2\pi I\over N}, {\rm for ~a~ 
closed ~chain} \nonumber \\
\end{eqnarray}
where the momentum $p$ is determined by an integer $I=1,2,...N$. The one-magnon 
eigenvalue is given by $\epsilon_1(p)=\epsilon_0-2J\cos{p}$.
The interaction strength $J$ determines the hopping of the down spins to 
neighbouring sites, and the interaction between the two down spins is determined 
by the anisotropy constant $\Delta$. The one-magnon eigen energies are 
independent of $\Delta$ as the states carry only one down spin. 

We will investigate the dynamics of quantum correlations by focusing on the 
evolution of a locally entangled state. For investigating the effect of a  
many-body interaction term in the Hamiltonian, we will consider two different 
initial states, a one-magnon entangled state and an entangled state in the zero 
and two-magnon subspace. The one-magnon eigenstates are not affected by the many 
body interaction term in the Hamiltonian, but the pair entanglement spreads out 
as the down spin moves around. The anisotropy parameter $\Delta$ plays a role in 
the case of the two-magnon state. 

Let us first consider an initial state that is a linear combination of localized 
one-magnon states, given by

  \begin{equation}
\vert \Psi(0)\rangle = \alpha \vert 100...0\rangle + \beta \vert 010...0\rangle 
= \alpha \vert 1\rangle+\beta\vert 2\rangle.
\label{initial_state_1}
 \end{equation}
 Here, we have written the initial state as a superposition of states with the 
down spin is at sites $1$ and $2$. 
 Through the time evolution the down spin moves due to the interaction. The time 
evolution of the state is straightforward, the state after a time $t$ becomes,
   \begin{equation}
\vert \Psi(t)\rangle= \sum_{x}\Omega^{x}(t)\vert x\rangle,
\label{heisenberg_state}
 \end{equation}
 where the time dependent function $\Omega^{x}(t)$ is given in terms of one 
particle Green functions \cite{subra1,saikat1} in the following form,
  
     \begin{equation}
\Omega^{x}(t) = \alpha G^{x}_{1}(t) +\beta G^{x}_{2}(t).
\label{green}
 \end{equation}
The time-dependent function $G^{x'}_{x}(t)$  \cite{subra1} is given in terms of 
the wave functions defined above as,  
   \begin{equation} 
 G^{x'}_{x}(t)=\sum_{p}\psi_{p}^x  \psi_p^{x'*} e^{-it\epsilon_1 (p)}.
  \end{equation}   
  In the macroscopic limit, $N\rightarrow\infty$, the sum over the momentum  can 
be converted  into an integral, thus, we get,
  \begin{equation}
  G^{x'}_x(t)={1\over 2 \pi}\int^{2\pi}_{0} e^{-it\cos{p}-ip(x-x')}dp= 
J_{x-x'}(t)i^{x-x'}.
  \label{bessel_function}
  \end{equation} 
 
  The reduced density matrix for the $j$th qubit can be worked out easily, we have
\begin{equation}
\rho_{j} = (1 - \vert \Omega^j\vert ^2)\vert 0\rangle \langle 0\vert  + \vert 
\Omega^j\vert^2 \vert 1\rangle \langle 1\vert.
   \label{eq:rdm1}
\end{equation} 
 The elements of the  two-qubit reduced density matrix (as shown in  
Eq.~\ref{two_party_rdm}), for the $j$th and $k$th qubits, for direct evolution 
of the states can be calculated from Eq.~\ref{heisenberg_state} and 
Eq.~\ref{green},

\begin{equation}
 u_{j,k}= 1-\vert \Omega^j\vert ^2-\vert \Omega^k \vert^2 , ~
 {w_1}_{j,k} = \vert \Omega^k\vert^2,~
 {w_2}_{j,k} = \vert\Omega^j\vert^2, ~
 x_{j,k} = \Omega^{j*}\Omega^k .
    \label{eq:rdm2}
  \end{equation}
The other elements of the reduced density matrix are zero for this case. 
Similarly, one can compute the elements of the three qubit reduced density 
matrix also. The forms of the elements are cumbersome to present here, so 
instead we directly present the expression for the TMI for three qubits. For the three marked qubits  $j, k$ and $l$, TMI can be 
expressed in terms of the two-body correlation functions given in 
Eq.~\ref{eq:rdm2}, and the Shannon binary entropy $H(x)=-x \log_2 x -(1-x) 
\log_2 (1-x)$, as

\begin{eqnarray}
I(j:k:l)= \sum_{p=j,k,l} H~( \vert {\Omega}^p \vert^2) + \sum_{p,q=j,k,l} H~ 
(\vert \Omega^p\vert^2 + \vert \Omega^q \vert^2) + H~(\vert\Omega^j\vert^2 + 
\vert \Omega^k\vert^2 +\vert\Omega^l\vert^2). \nonumber\\
\label{eq:tmi1}
\end{eqnarray}
 Here, the quantities $\vert \Omega^j \vert^2, \vert \Omega^k \vert^2,$ and 
$\vert \Omega^l \vert^2$ are clearly non negative fractions, implying that 
$I(j:k:l)$ is non negative, as shown in Fig.~\ref{fig:3}(a). Thus, the 
scrambling of information does not occur in this case.


Let us now consider an initial state that is an entangled state in zero and two 
magnon subspace, given as

  \begin{equation}
\vert\Psi(0)\rangle =  \alpha\vert0...0\rangle+\beta\vert110...0\rangle = \alpha 
\vert F \rangle+\beta \vert 1,2\rangle.
 \end{equation}
 The time evolution of the state is straightforward, the state after a time $t$ 
becomes,
   \begin{equation}
\vert \Psi(t)\rangle= \alpha e^{-i\epsilon_0 t} \vert F\rangle + 
\sum_{x_1,x_2}G^{x_1,x_2}_{1,2}(t) \vert x_1,x_2\rangle,
\label{state_12_t}
 \end{equation}
The details of the two-particle time-dependent Green function 
$G^{x_1,x_2}_{1,2}(t)$ is discussed in \cite{saikat2}.  The two-particle Green's 
function is defined in terms of the two-magnon eigenfunctions $\psi_{p_1,p_2} 
(x_1,x_2)$,  and the eigenvalues $\epsilon_2 (p_1,p_2)$,  we have,
   \begin{equation} 
 G^{x'_1,x'_2}_{x_1,x_2}(t)=\sum_{p_1,p_2}\psi_{p_1,p_2}^{x_1,x_2} 
\psi_{p_1,p_2}^{x'_1,x'_2 *} e^{-it\epsilon_2 (p_1,p_2)}.
  \end{equation} 
 The  elements of the one and two-qubit reduced density matrix for direct 
evolution of the states can be calculated from Eq.~\ref{state_12_t}.  The 
reduced density matrix for the $j$th qubit is given as,
 \begin{equation}
\rho _{j} = (1 - \vert \beta \vert^2 \sum_{x; x \neq j} \vert 
G^{j,x}_{1,2}\vert^2)\vert 0\rangle \langle 0\vert +\vert \beta \vert^2 \sum_{x; 
x \neq j} \vert G^{j,x}_{1,2}\vert^2 \vert 1\rangle \langle 1\vert.
\end{equation} 
The elements of the reduced density matrix $\rho_{j,k}$ given in 
Eq.~\ref{two_party_rdm} are given as,

\begin{eqnarray}
  &&u_{j,k}= \vert\alpha\vert^2 + \vert \beta \vert^2 \sum_{x_1,x_2; x_1,x_2 
\neq j,k} \vert G^{x_1,x_2}_{1,2} \vert^2, \nonumber\\
 &&{w_1}_{j,k} = \vert \beta \vert^2 \sum_{x; x \neq j} \vert 
G^{k,x}_{1,2}\vert^2,\nonumber\\
  &&{w_2}_{j,k} = \vert \beta \vert^2 \sum_{x; x \neq k} \vert 
G^{j,x}_{1,2}\vert^2, ~~~
  x_{j,k} = \vert \beta \vert^2 \sum_{x; x \neq j,k} 
G^{j,x}_{1,2}G^{*k,x}_{1,2}, \nonumber\\
  &&v_{j,k} = \vert \alpha \vert^2, ~~~
  z_{j,k} = \alpha \beta^*e^{-i\epsilon_0 t} G^{*j,k}_{1,2}.
\end{eqnarray}

The two particle Green functions $G^{x_1,x_2}_{1,2}$ have been calculated 
numerically for different values of anisotropy constant $\Delta$. Accordingly, 
the three-qubit reduced density matrix $\rho_{j,k,l}$ for three qubits $j$, $k$ 
and $l$  has been calculated in order to compute the TMI. We illustrate the 
results for the initial states with a Bell pair at the first two spins i.e., 
$(\vert10...0\rangle+\vert010...0\rangle)/\sqrt{2}$ and 
$(\vert00...0\rangle+\vert110...0\rangle)/\sqrt{2}$. The nearest neighbour 
concurrence $C(i,i+1)$, mutual information $I(i,i+1)$ are plotted as functions 
of time $t$ and site index $i$ in Fig.~\ref{fig:2}(a) and \ref{fig:2}(b) for the 
initial  state $\vert\Psi(0)\rangle = 
\frac{1}{\sqrt{2}}(\vert10...0\rangle+\vert010...0\rangle)$ and  in  Fig.~ 
\ref{fig:2}(c) and \ref{fig:2}(d) for the initial  state $\vert\Psi(0)\rangle = 
\frac{1}{\sqrt{2}}(\vert00...0\rangle+\vert110...0\rangle)$ respectively for the 
anisotropic Heisenberg model. Both the concurrence and the mutual information are 
monotonic and show similar features. The pairwise entanglement and the 
correlations moves linearly in time as time evolves from the first two sites to 
the rest of the chain.\\

Tripartite mutual information  $I(1:2:3)$ is plotted as a function of time $t$ 
for the initial state $(\vert100...0\rangle+\vert010...0\rangle)/\sqrt{2}$ in 
Fig.~\ref{fig:3}(a) and for different values of the anisotropic parameter 
$\Delta$  in Fig.~\ref{fig:3}(b) respectively. Unlike for the initial state in 
one magnon subspace,  we see here that the matrix element $z$ of two-qubit 
reduced density matrix is non zero indicating the mixing between the zero and 
the two magnon sectors. As a result, the TMI can be negative despite the 
dynamics being spin-conserving. The scrambling is very small as the minimum 
value of TMI is around $-0.2$.

 \begin{figure*}[t]
     \begin{center}
        \subfigure[]{%
            \includegraphics[width=0.35\textwidth]{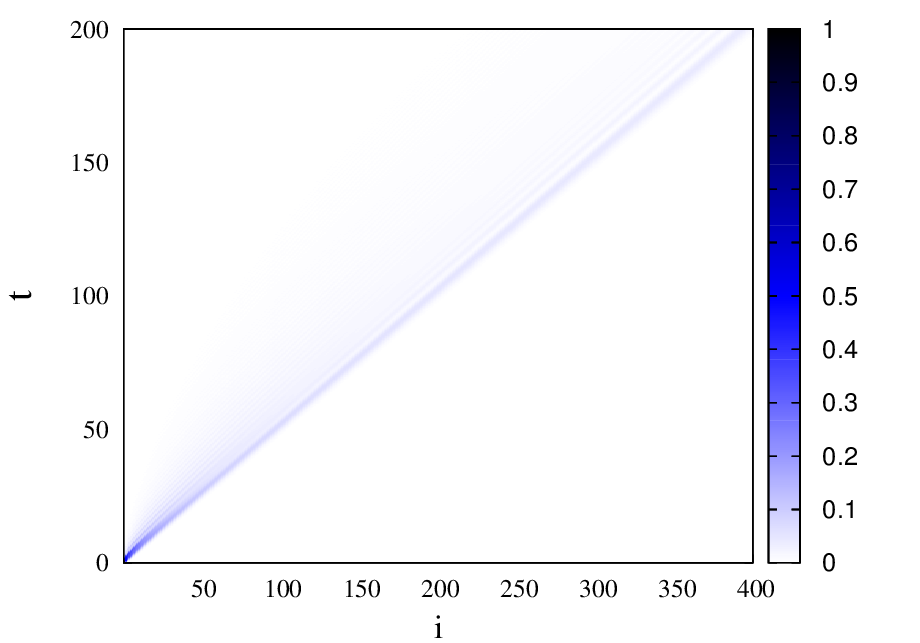}
        }%
        \subfigure[]{%
           \includegraphics[width=0.35\textwidth]{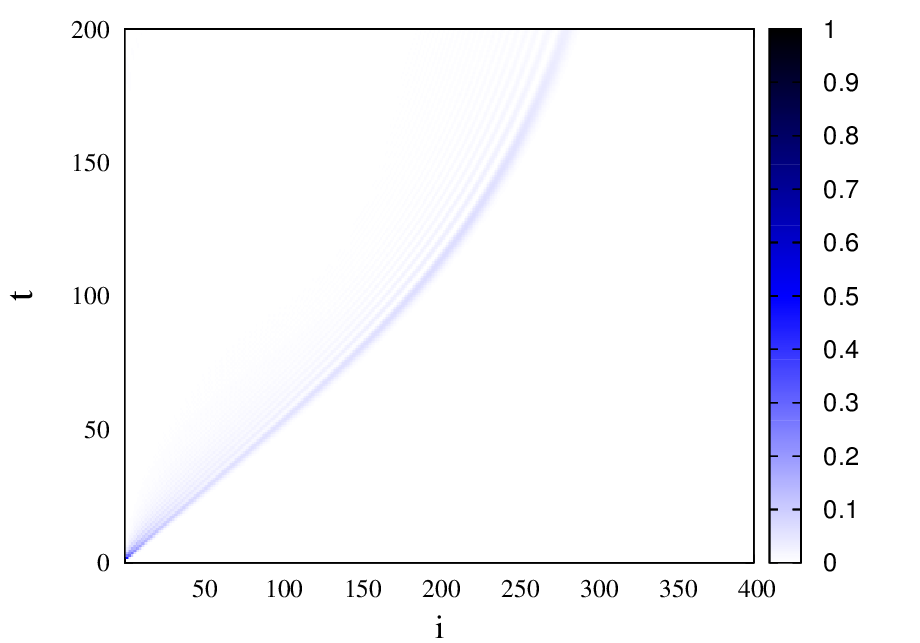}
        }\\%
        \subfigure[]{%
            \includegraphics[width=0.35\textwidth]{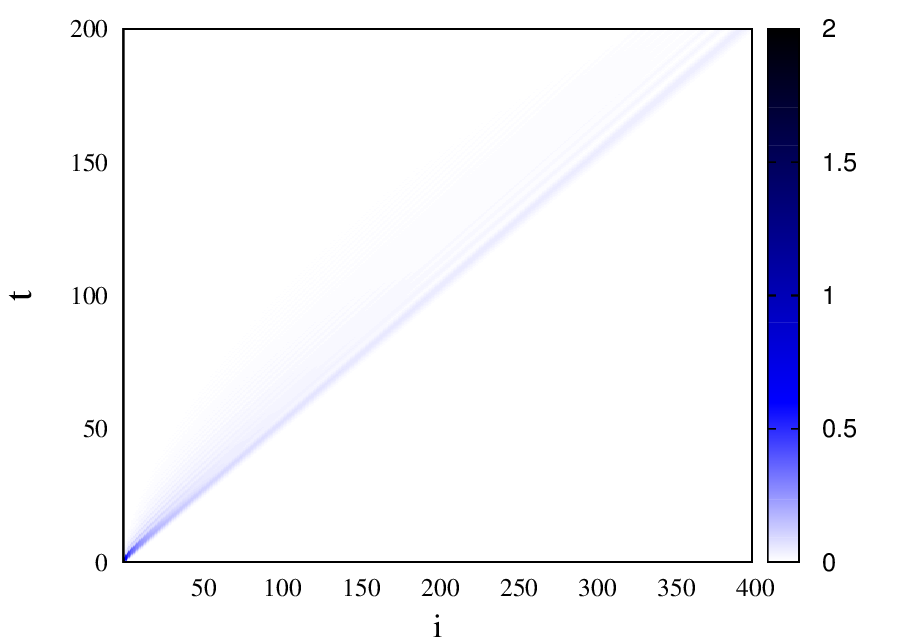}
        }%
        \subfigure[]{%
           \includegraphics[width=0.35\textwidth]{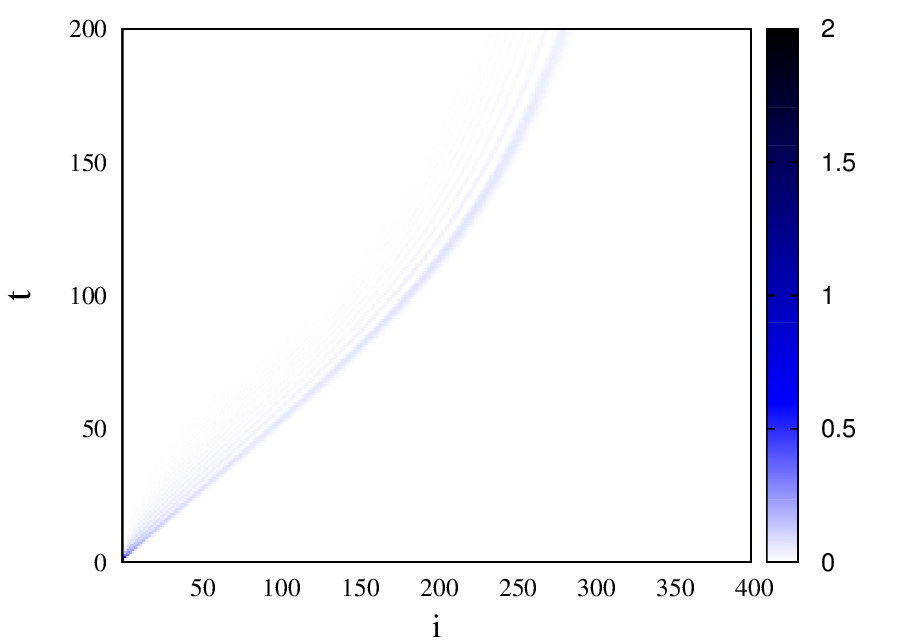}
        }%
    \end{center}
\caption{{The time dependence of the concurrence and the mutual information  
between the sites $i$ and $i+1$ for the Harper model: The concurrence is shown 
for parameters (a) $g = 0.1, \tau = 0.1$, (b)  $g = 1.0, \tau = 0.9$. The mutual 
Information is shown for parameters (c) $g = 0.1, \tau = 0.1$, (d)  $g = 1.0, 
\tau = 0.9$. The results are shown for the initial state 
$\vert10...0+010...0\rangle/\sqrt{2}$.}
     }%
   \label{fig:4}
   \label{figure5}  
\end{figure*}
\begin{figure}[h]
\begin{center}
        \subfigure[]{%
            \includegraphics[width=0.35\textwidth]{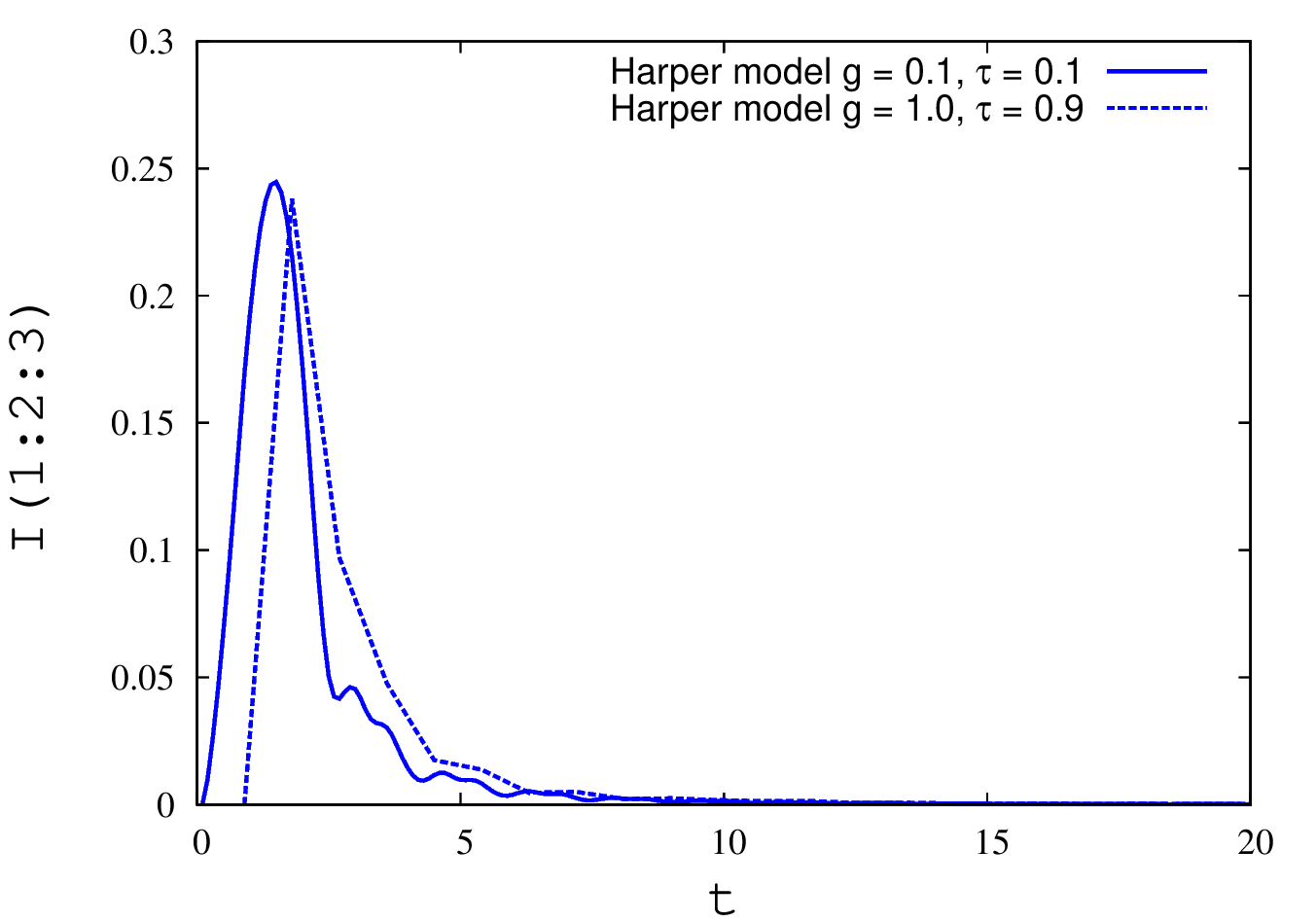}
        }%
        \subfigure[]{%
           \includegraphics[width=0.35\textwidth]{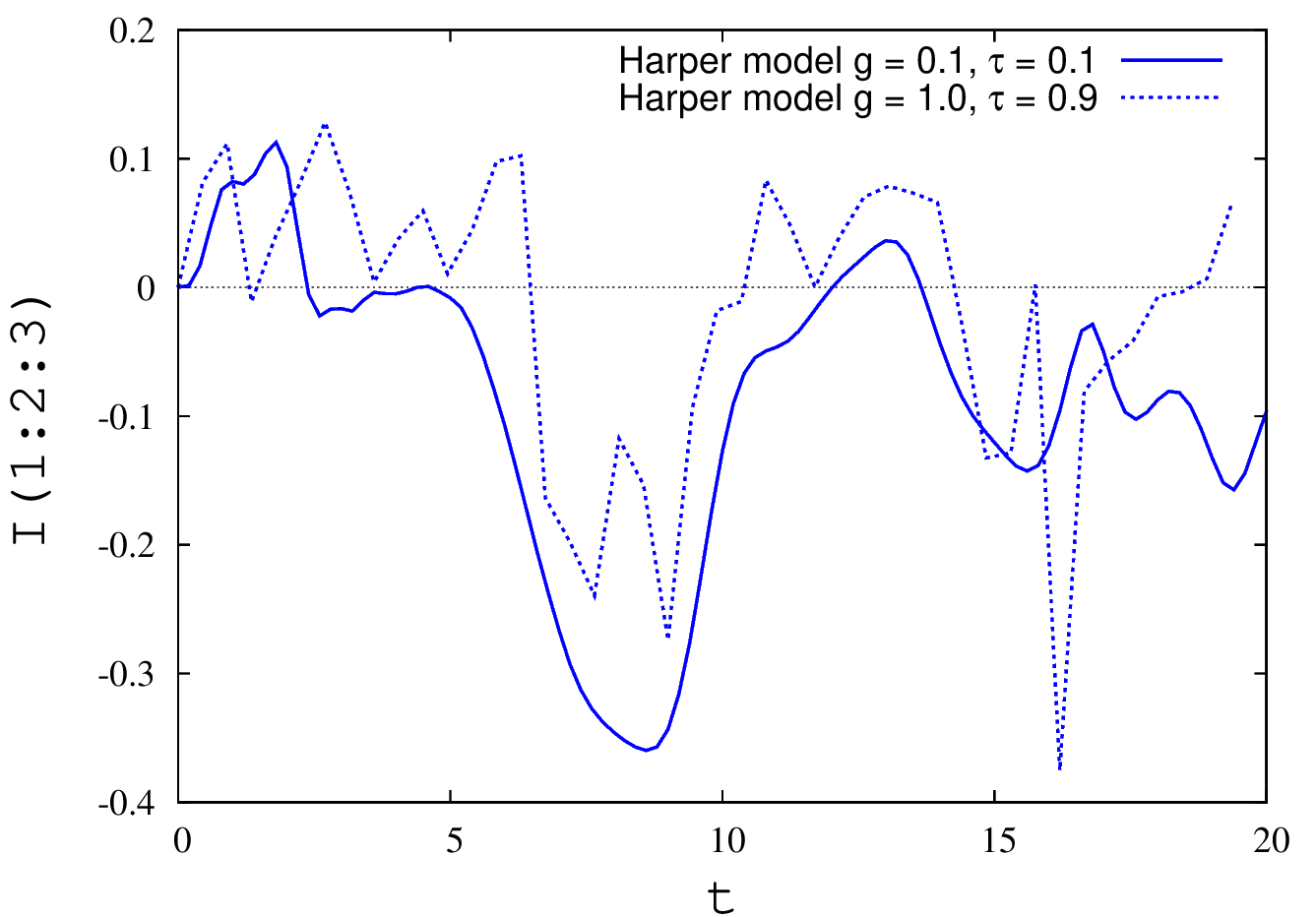}
        }\\%
\end{center}
\caption{{The time dependence of TMI $I(1:2:3)$ for the Kicked Harper model with 
 different set of parameters $\tau = 0.1 , g = 0.1$ and $\tau = 0.9 , g = 1.0$ 
as a function of time $t$. }}
\label{fig:5}
\end{figure}

 \subsection{Periodically Kicked Harper model}
 Next, we consider a simple model Hamiltonian with a tunable parameter, to go 
continuously from a completely integrable to a completely non-integrable regime. 
We use a one-dimensional periodically-kicked Harper model, a simple model of 
fermions hopping on a chain with an inhomogeneous site potential, appearing as a 
kick at regular intervals. The spin operator version of the Hamiltonian is given 
by
 \begin{eqnarray} 
H(t) &&= \sum_{j=1}^N 
[-\frac{1}{2}({\sigma}^x_{j}{\sigma}^x_{j+1}+{\sigma}^y_{j}{\sigma}^y_{j+1})  +g 
\sum_{n=- \infty}^{\infty} \delta(\frac{t}{\tau}-n)\cos(\frac{2\pi j \eta}{N}) 
{\sigma}^z_{j}].
  \end{eqnarray} 
  The first term is the XY term of the Heisenberg model considered above, that 
causes hopping of up or down spins. The last term is an inhomogeneous magnetic 
field in the transverse direction that   comes into play through kicks at an 
interval of $\tau$. The coupling strength $g$ and the kicking interval $\tau$ 
can be independently varied, that can affect the nature of the dynamics as we 
will see below. 
 
The classical version of the kicked Harper Hamiltonian is regular for $\tau 
\rightarrow 0$ and completely chaotic for large values of $\tau$, and similarly, 
the eigenvalues and the eigenfunctions of the quantum version display 
correspondingly a regular or a chaotic characteristics~\cite{lima, arul_2003, 
sur_ghosh}. We  consider evolution at discrete  times, viz., $t=\tau^+,2\tau^+$ 
etc, i.e. at instants just after a kick. The unitary operator for the evolution 
between two kicks is straightforwardly given by,
 \begin{equation}
 U(g,\tau) = e^{-i \tau \sum_{j} 
-\frac{1}{2}({\sigma}^x_{j}{\sigma}^x_{j+1}+{\sigma}^y_{j}{\sigma}^y_{j+1})}  
e^{-i \tau g \sum_j \cos{\frac{2\pi j \eta}{N}} {\sigma}^z_{j} },
\end{equation}
where, the two operator factors appearing above do not commute. The time evolved 
state at time $n\tau^+$ just after $n$ kicks is 
$\vert\Psi(t)\rangle=U^n(g,\tau)\vert\Psi(0)\rangle$. The system evolves between 
two kicks at $n\tau^+$ and $(n+1)\tau^-$ through the XY dynamics but kicks 
introduce a lattice position dependent phase factor to the Green function. We 
consider the initial state $\vert\Psi(0)\rangle = \alpha \vert10...0\rangle + 
\beta \vert010...0\rangle$ of the system and the time evolved state will be 
given by,
 \begin{equation} 
\vert \tilde{{\Psi}}(t = n\tau^+)\rangle = \sum_{x}\tilde{\Omega}^{x}(t = 
n\tau)\vert x\rangle,
  \end{equation}
  
  where, $\tilde{\Omega}^{x}(t = n\tau) =   \sum_{x} \alpha \tilde{G}^x_1(t = 
n\tau)\vert x\rangle +\beta\tilde{G}^x_2(t = n\tau)\vert x\rangle$.
  
 Here we have introduced a composite Green function~\cite{saikat2,sur_ghosh}, 
related to the Green function studied in the Heisenberg dynamics, given in the 
following form,
  \begin{equation} 
\tilde{G}^{x_{n}}_{x_0}(t = n\tau) = \sum_{x_1,x_2,...,x_n} 
\prod^{n-1}_{j=0}{G}^{x_{j+1}}_{x_j}(\tau)e^{2i\tau{ g\cos(\frac{2\pi \eta 
x_{j+1}}{N})}}.
\label{composite_green_function}
  \end{equation}  
It can be seen that after each kick, a site-dependent new phase is introduced in 
the Green function which indicates a qualitative change in the dynamics from  
the previously discussed Heisenberg dynamics. By setting  $g\tau = 0$ in the 
above, the composite Green function $\tilde{G}^{x}_{1}(t)$  reduces to the Green 
function ${G}^{x}_{1}(t)$, the one-magnon Green function of the Heisenberg 
model. The form of the reduced density matrices will remain unchanged as given 
in Eq.~\ref{eq:rdm1} and Eq.~\ref{eq:rdm2}.  Let us first consider the initial 
state given in Eq.~\ref{initial_state_1}, a linear combination of one magnon 
states. Since the system here is non-interacting, the time-dependent wave 
function from any initial state can be written as a product of the Green 
function given in Eq.~\ref{composite_green_function}.  The form of the two qubit 
RDM derived for the Heisenberg dynamics with one magnon states in 
Eq.~\ref{eq:rdm2} is also valid for the Harper dynamics with the composite Green 
function.\\

The nearest neighbour concurrence $C(i,i+1)$ and the mutual information 
$I(i,i+1)$ are plotted for two representative values of $\tau$ and $g$ as 
functions of site index $i$  and time $t$ for the kicked Harper model in 
Fig.~\ref{fig:4}(a)- \ref{fig:4}(b). The pairwise concurrence and the mutual 
information moves from the first pair to large distances with time. For $\tau = 
0.1$ and $g = 0.1$ the dynamics resembles the Heisenberg dynamics, where 
correlations spread linearly as seen in Fig.~\ref{fig:4}(a) and \ref{fig:4}(c). 
But for $\tau = 0.9$ and $g = 1.0$, the light cone structure becomes non linear 
as seen in Fig.~\ref{fig:4}(b) and \ref{fig:4}(d). Even in this regime, the 
dynamics is not much different from the Heisenberg dynamics as the number of 
magnons is conserved. The correlation dynamics does not change qualitatively for 
large values of $g\tau$, where the dynamics is nonintegrable. The TMI of  the first 
three qubits $I(1:2:3)$ has been plotted for the kicked Harper model for both 
the cases $\tau = 0.1, g = 0.1$  and $\tau = 0.9, g = 1.0$ for the initial state 
$\vert\Psi(0)\rangle = (\vert100...0\rangle+\vert010...0\rangle)/\sqrt{2}$.  As 
shown in Fig.~\ref{fig:5}(a), the quantity $I(1:2:3)$ is non negative for all 
the cases. One can also argue this from Eq.~\ref{eq:tmi1} that TMI is 
non-negative when the dynamics is spin-conserving and confined to a one magnon 
sector. For the initial state  $\vert\Psi(0)\rangle = 
(\vert000...0\rangle+\vert110...0\rangle)/\sqrt{2}$, the quantity $I(1:2:3)$ can 
be negative, as shown in  Fig.~\ref{fig:5}(b). Similar to the Heisenberg 
dynamics, the zero and the two magnon sectors mix as the matrix element $z$ of 
the two-qubit reduced density matrix is non zero.

 Till now we have seen that the qualitative nature of spreading of correlations 
from a maximally entangled pair or Bell pair does not depend on the 
integrability of the dynamics. But confinement of  the dynamics into a much 
smaller subspace of the Hilbert space leads to better transfer of quantum 
correlations in a many body system. As  we have seen that bipartite correlations 
spread to long distances for the Heisenberg model and the kicked Harper model, 
where the number of down (up) spins is a conserved quantity. So, it can be 
concluded that spreading of bipartite quantum correlations in a quantum 
many-body system is associated with a non-negative value of TMI as in the cases 
of the Heisenberg and the kicked Harper model.

   \begin{figure*}[t]
     \begin{center}

        \subfigure[]{%
            \includegraphics[width=0.28\textwidth]{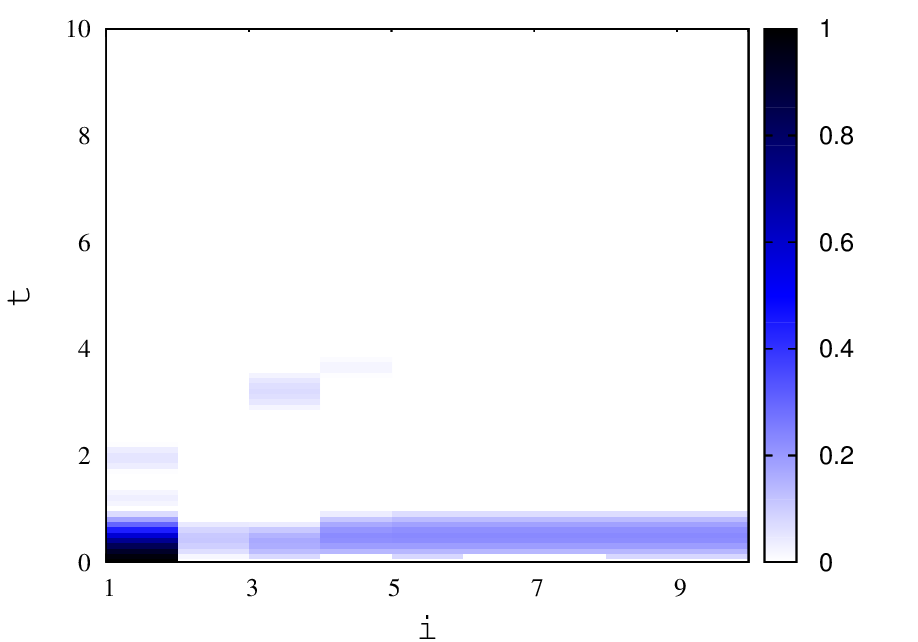}
        }%
        \subfigure[]{%
           \includegraphics[width=0.28\textwidth]{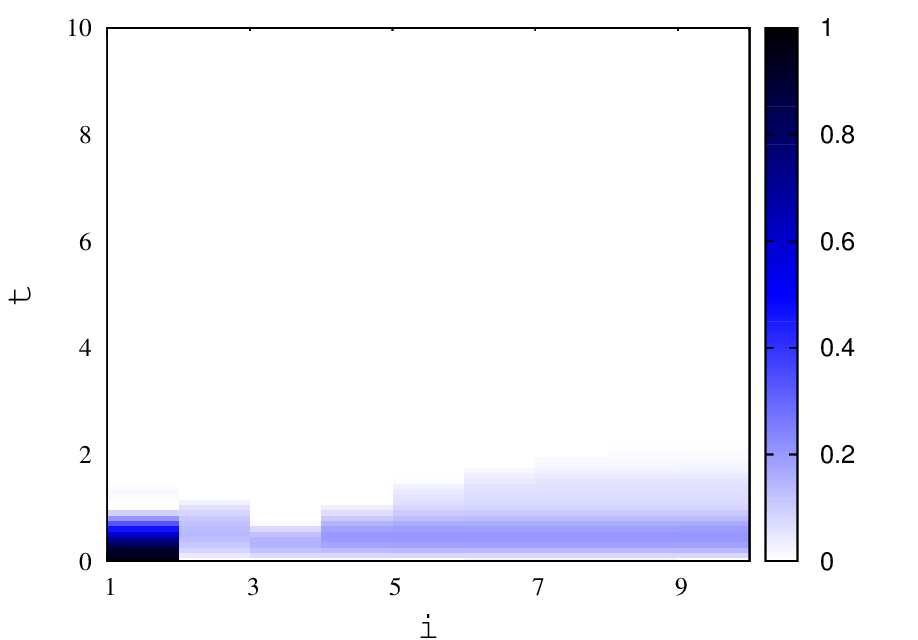}
        }%
        \subfigure[]{%
            \includegraphics[width=0.28\textwidth]{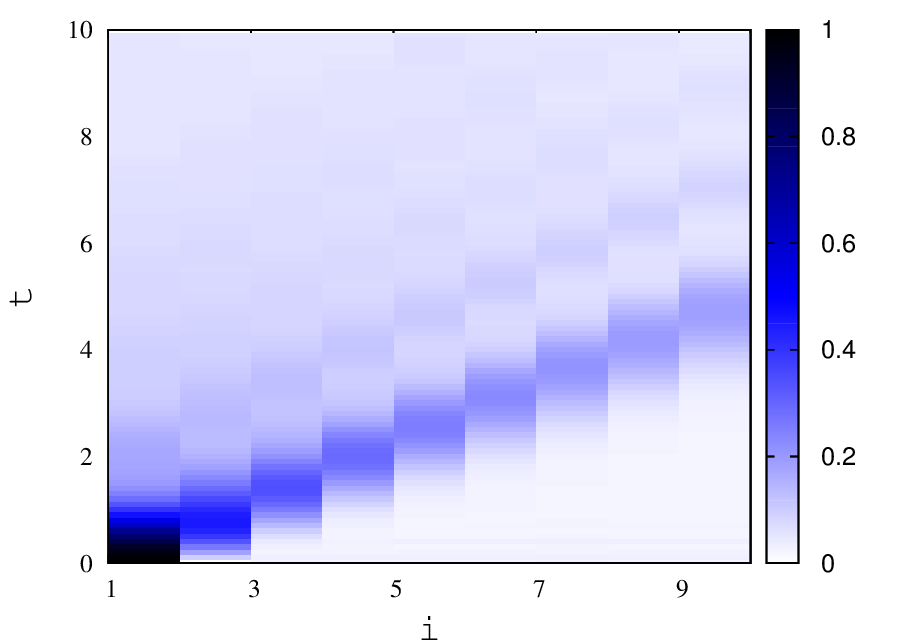}
        }\\%
        \subfigure[]{%
           \includegraphics[width=0.28\textwidth]{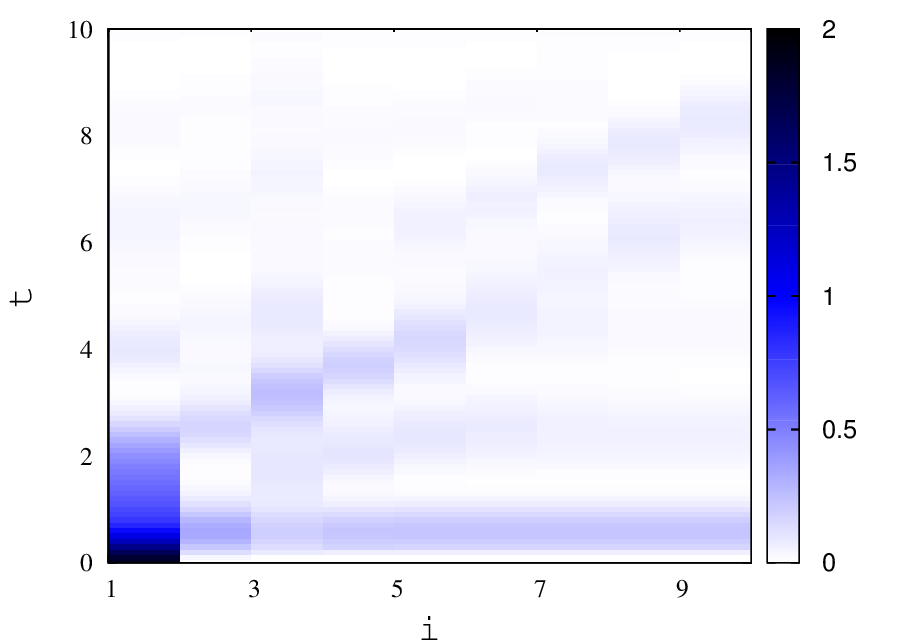}
        }%
        \subfigure[]{%
           \includegraphics[width=0.28\textwidth]{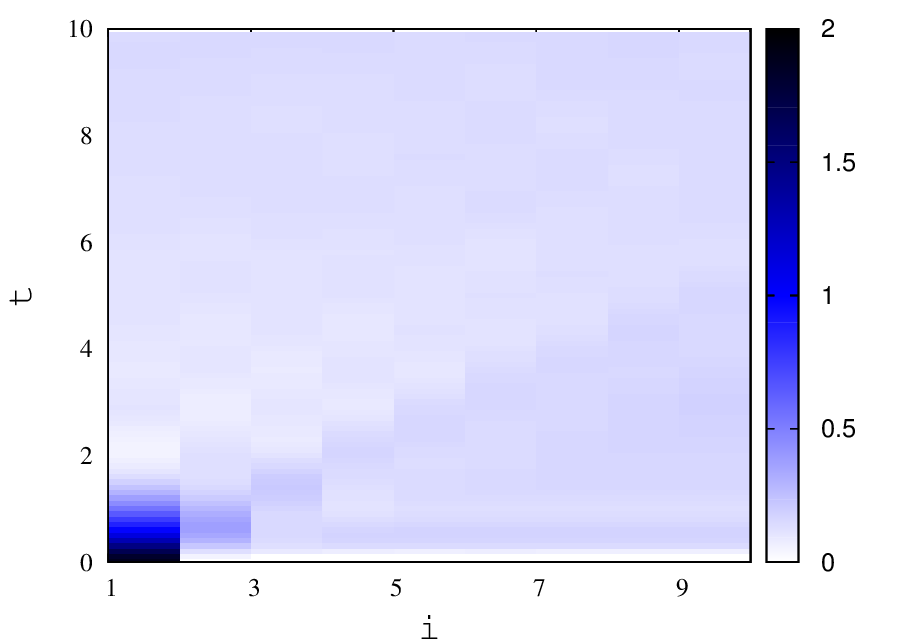}
        }%
        \subfigure[]{%
           \includegraphics[width=0.28\textwidth]{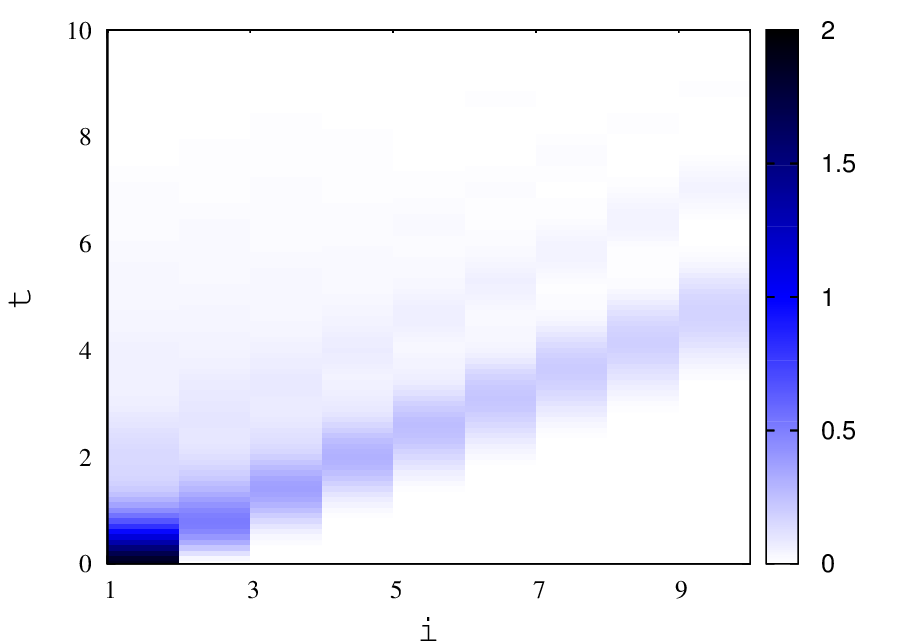}
        }\\%
    \end{center}
\caption{{The concurrence and the mutual information measures between 
the sites $i$ and $i+1$ as functions of time $t$ for the XY model with a 
transverse field:  The concurrence shown  for parameters: (a) $J_x =0.7, J_y = 
0.3$ and $h = 0.1$, (b) $J_x =0.7, J_y = 0.3$ and $h = 1.0$, (c) $J_x =0.7, J_y 
= 0.3$ and $h = 10.0$. The mutual Information for parameters (d) $J_x =0.7, J_y 
= 0.3$ and $h = 0.1$, (e) $J_x =0.7, J_y = 0.3$ and $h = 1.0$, (f) $J_x =0.7, 
J_y = 0.3$ and $h = 10.0$. The results are shown from analytical calculations 
for the initial state:  $\vert\Psi(0)\rangle = (\vert10...0\rangle+\vert010...0\rangle)/\sqrt{2}$.}
    }%
  \label{fig:6}  
\end{figure*}

 \begin{figure*}[t]
     \begin{center}

        \subfigure[]{%
           \includegraphics[width=0.35\textwidth]{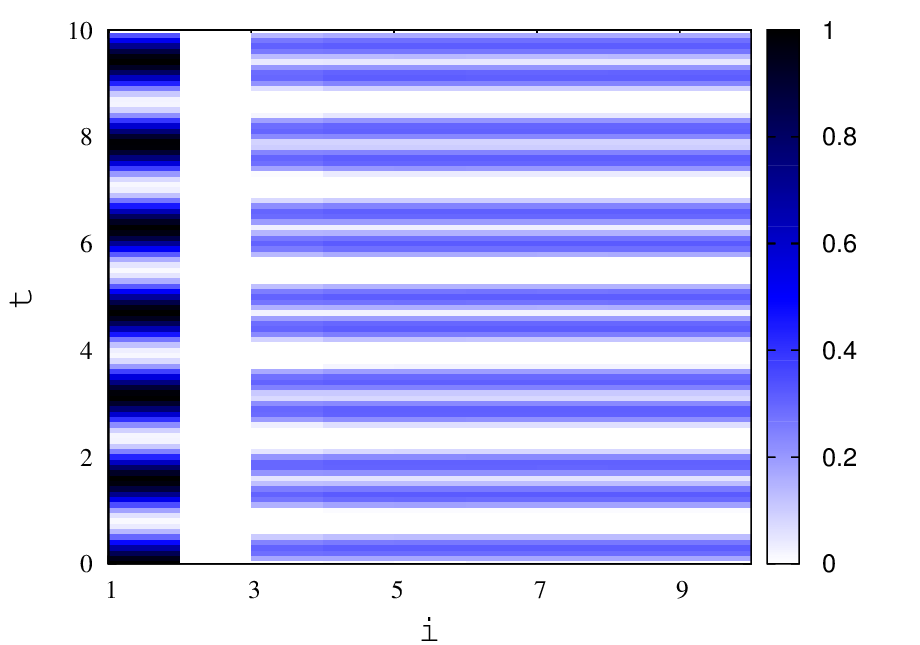}
       }%
       \subfigure[]{%
          \includegraphics[width=0.35\textwidth]{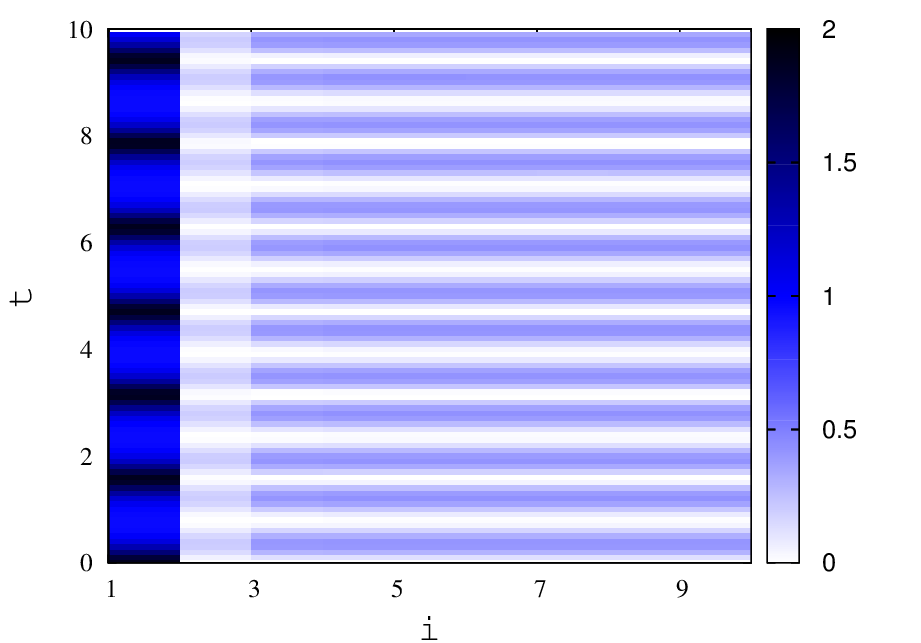}
       }\\%
  \end{center}
\caption{{The concurrence and the mutual information measures are shown between 
the sites $i$ and $i+1$ as  functions of time $t$ for the Ising model: (a) the 
concurrence, (b) the mutual information. The results are shown from analytical 
calculations for the initial state $ \vert10...0+010...0\rangle/\sqrt{2}$.}
     }%
   \label{fig:7}  
\end{figure*}

 \begin{figure*}[t]
     \begin{center}

        \subfigure[]{%
         {~~~}  \includegraphics[width=0.35\textwidth]{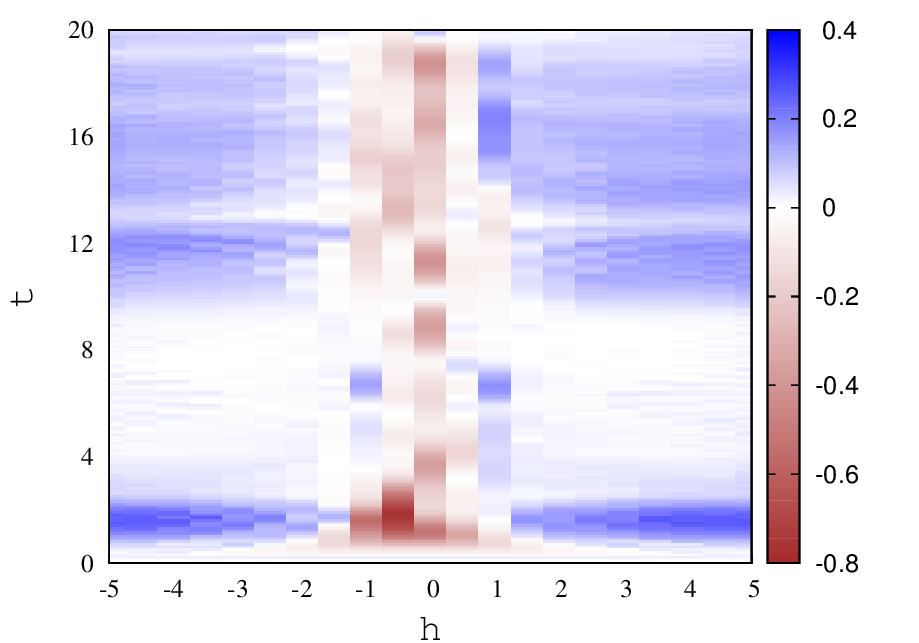}
       }%
       \subfigure[]{%
          \includegraphics[width=0.35\textwidth]{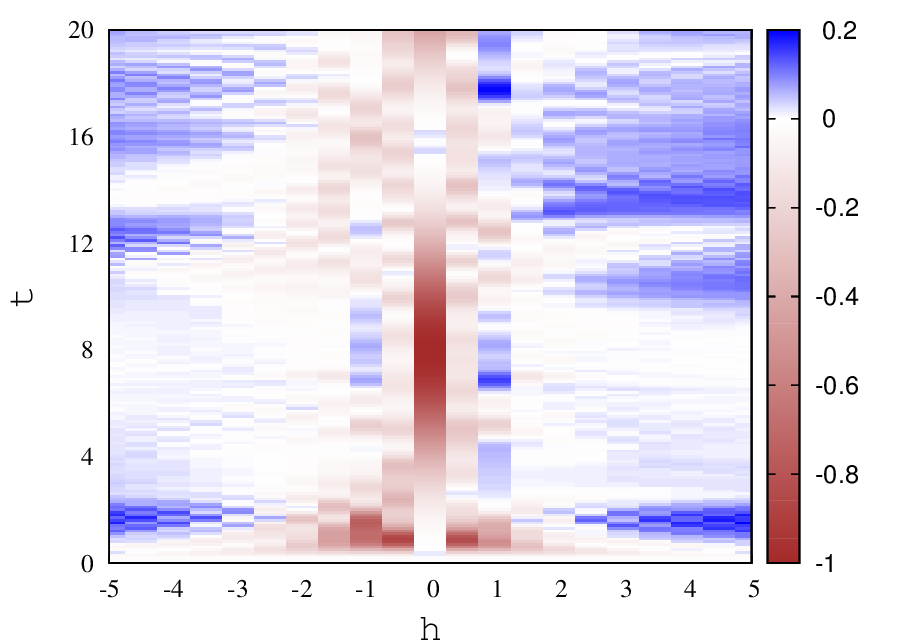}
       }\\%
               \subfigure[]{%
           {~~} \includegraphics[width=0.33\textwidth]{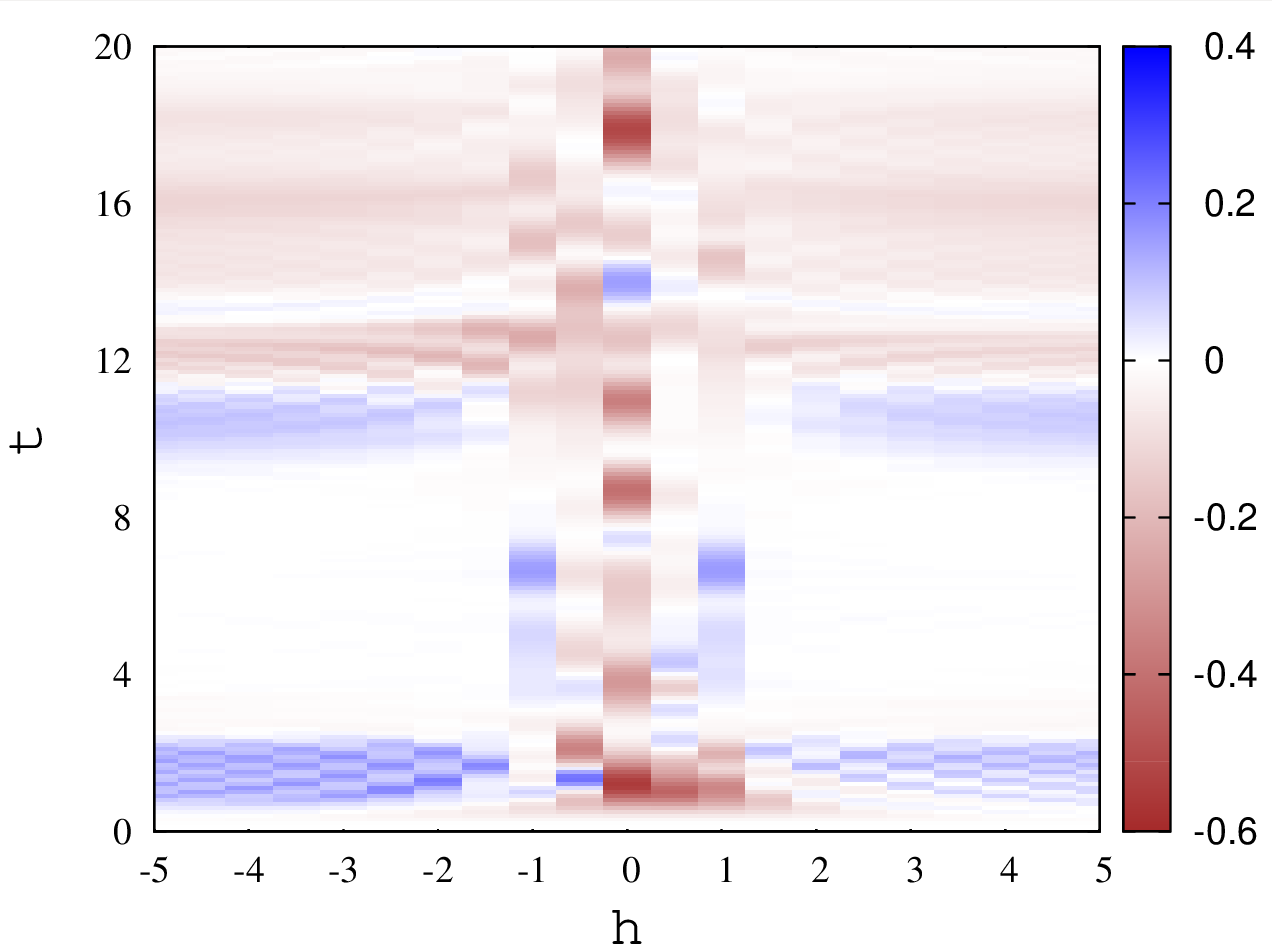}
       }%
       \subfigure[]{%
          {~} \includegraphics[width=0.33\textwidth]{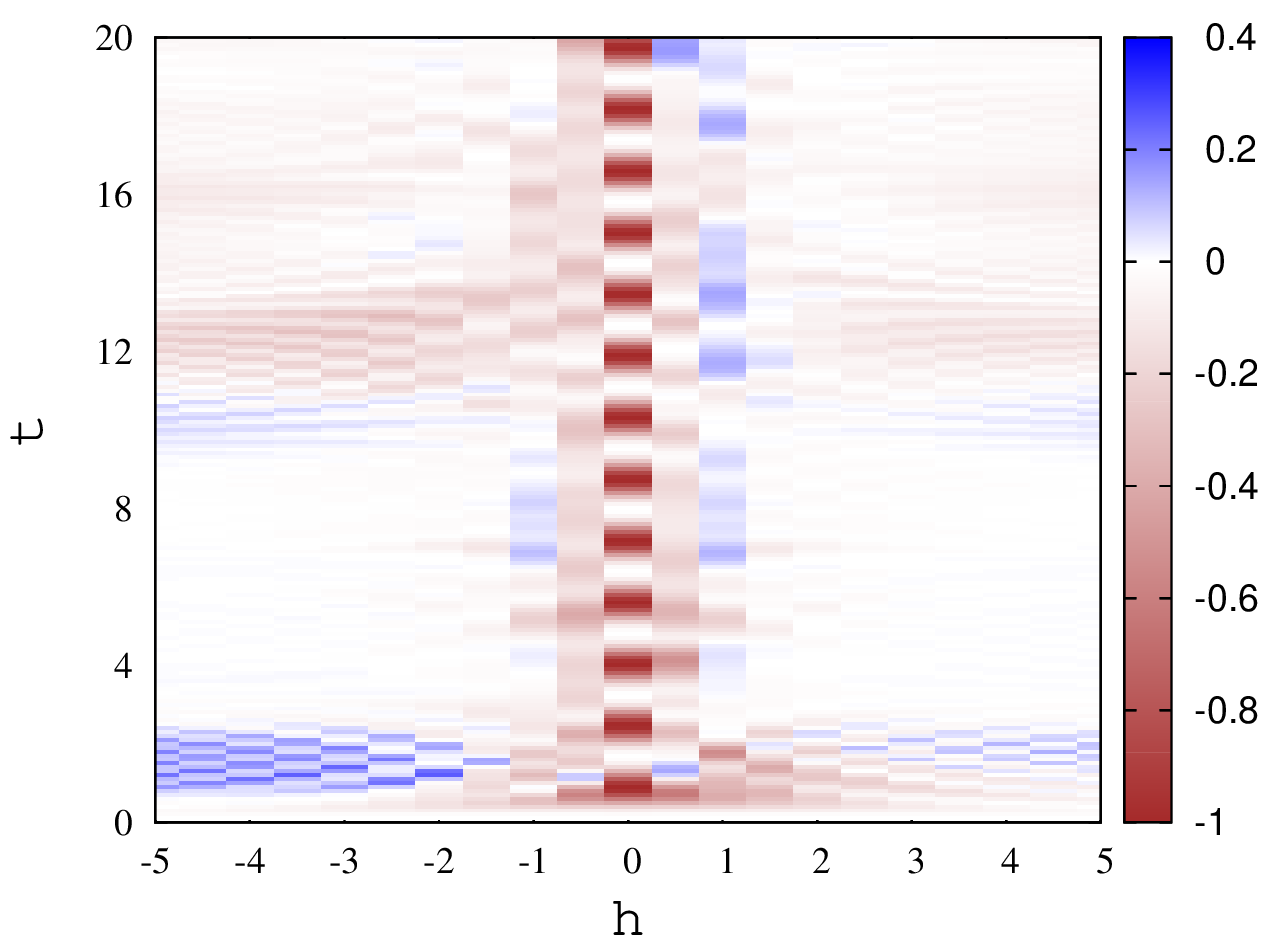}
       }\\%
  \end{center}
\caption{{The time dependence of $I(1:2:3)$ with time: (a)  for the XY model 
($J_x = 0.7$, $J_y = 0.3$) with different values of transverse magnetic field 
$h$, (b) for the Ising model with different values of transverse magnetic field 
$h$ for the initial state $\vert10...0+010...0\rangle/\sqrt{2}$; (c)  for the XY 
model ($J_x = 0.7$, $J_y = 0.3$) with different values of transverse magnetic 
field $h$, (d) for the Ising model with different values of transverse magnetic 
field $h$ for the initial state $\vert00...0+110...0\rangle/\sqrt{2}$.}
     }%
   \label{fig:8}  
\end{figure*}

\subsection{Transverse field XY model}
Till now, we have considered model Hamiltonians that conserve the total 
magnetization. Now, we consider the XY model with a transverse magnetic field, 
where the dynamics is spin non-conserving. The general Hamiltonian for a one 
dimensional XY model with $N$ spins with a magnetic field in the transverse 
direction is given as,
 
   \begin{equation}
H = \sum_{i}J_x{\sigma}^x_{i}{\sigma}^x_{i+1}+J_y{\sigma}^y_{i}{\sigma}^y_{i+1} 
+ h{\sigma}^z_{i}.
\end{equation} 

Here, the periodic boundary condition is assumed. It is easy to see that the 
three different terms in the above Hamiltonian do not commute with each other 
for $J_x \neq J_y \neq 0$. The ground state  exhibits a quantum critical 
behaviour, for the isotropic case of $J_x=J_y$ for all values of the magnetic 
field strength, and for the anisotropic case for  $h=J_x+J_y$. Since the spin 
$1/2$ operators are neither bosons nor fermions, the Hamiltonian can be exactly 
diagonalized and the entire eigenvalue spectrum can be found by employing 
Jordan-Wigner transformation~\cite{jordan, lieb} that maps spin $1/2$ operators 
to spinless fermionic operators. The mapping is given by,
 \begin{equation}
\sigma_l^{+} = c^{\dagger}_le^{i\pi \sum_{m=1}^{l-1} c_m^{\dagger}c_m }.
 \end{equation}
 The Hamiltonian, having a bilinear form in terms of Fermionic creation and 
annihilation operators, can be  brought to a diagonal form by doing a Fourier 
transformation, followed  by a Bogoliubov-Valatin transformation 
\cite{bogoliubov, valatin}. Fourier transformation maps the operators into 
momentum space, as follows 
\begin{equation}
c_q={1\over \sqrt{N}}\sum e^{-iql}c_l.
\end{equation}
Here, the set of allowed momentum values are given by  $q=2\pi m/N$, with $ m = 
-(N-1)/2,...,-1/2,1/2,...,(N-1)/2$ for even values of $N$; and $m = 
N/2,...,0,...,N/2$ for odd values of $N$. In terms of these momentum-space 
operators the Hamiltonian has a bilinear form with non-diagonal operators 
$c^{\dagger}_qc^{\dagger}_{-q}$ and similar terms.

To diagonalize the Hamiltonian we employ Bogoliubov-Valatin transformation in 
which new Fermion creation and annihilation operators  
are formed as a linear combination of old operators, 
\begin{equation}
\eta_{1q}=u_q c_{q}-i v_q c^\dagger_{-q},{~ ~} \eta_{2q}=-iv_q c_q +u_q 
c^{\dagger}_{-q}.
\end{equation} The expansion coefficients and the eigenvalues are given by the 
following forms,
\begin{equation}
u_q = \sqrt{{1\over 2}+{(J_x+J_y)\cos q + h\over \vert\omega_q\vert}},~~v_q = 
\sqrt{1-u_q^2},
\end{equation} \begin{equation}
\omega_q = 2\sqrt{[(J_x+J_y) \cos q +h]^2+[(J_x-J_y)\sin q]^2}.
\end{equation}
In terms of these new fermion operators the Hamiltonian is diagonal and is given 
as,
\begin{equation}
H = \sum_{0<q<\pi} \vert\omega_q\vert (\eta^{\dagger}_{1q} \eta_{1q} - 
\eta^{\dagger}_{2q} \eta_{2q}).
\end{equation}

The elements of the reduced density matrix, given in Eq.~\ref{two_party_rdm}, in 
terms of spin operators and fermion operators are given by,
 
 \begin{eqnarray} 
u_{j,k} &&= \frac{1+ \langle\sigma_j^z\rangle + \langle\sigma_k^z\rangle 
+\langle\sigma_j^z\sigma_k^z\rangle}{4}  = {1-\langle c^{\dagger}_j c_j 
\rangle-\langle c^{\dagger}_k c_k \rangle}+ \langle c^{\dagger}_j c_j 
c^{\dagger}_k c_k \rangle,\nonumber\\
v_{j,k} &&=\frac{1- \langle\sigma_j^z\rangle - \langle\sigma_k^z\rangle + 
\langle\sigma_j^z\sigma_k^z\rangle}{4} = \langle c^{\dagger}_j c_j c^{\dagger}_k 
c_k \rangle,\nonumber\\
{w_1}_{j,k} &&= \frac{1- \langle\sigma_j^z\rangle + \langle\sigma_k^z\rangle 
-\langle\sigma_j^z\sigma_k^z\rangle}{4} =  1-\langle c^{\dagger}_k c_k \rangle - 
\langle c^{\dagger}_j c_j c^{\dagger}_k c_k \rangle,\nonumber\\
{w_2}_{j,k} &&=\frac{1+ \langle\sigma_j^z\rangle - \langle\sigma_k^z\rangle 
-\langle\sigma_j^z\sigma_k^z\rangle}{4} = 1-\langle c^{\dagger}_j c_j \rangle - 
\langle c^{\dagger}_j c_j c^{\dagger}_k c_k \rangle,\nonumber\\
x_{j,k} &&=  \langle\sigma_j^+\sigma_k^-\rangle = \langle c_j c^{\dagger}_{k} 
\rangle  \mbox{ for } k =j+1,\nonumber\\
z_{j,k} &&= \langle\sigma_j^+\sigma_k^+\rangle = \langle c_j c_{k} \rangle 
\mbox{ for } k =j+1.\nonumber\\
\label{correlation_fn}
 \end{eqnarray}
 
 Now, the time evolution of the fermion annihilation operator $c_q$ in the 
momentum space becomes,
\begin{eqnarray}
c_q(t) =  \chi_q(t)~c_q + \xi_q(t) ~c_{-q}^\dagger,
\end{eqnarray}
with time dependent functions $\chi_q(t) = (e^{-i\omega_qt}u_q^2+ e^{i\omega_qt} 
v_q^2)$ and $\xi_q(t) = - {q\over \vert q\vert} 2 u_q v_q \sin{\omega_q t}$.

 Let us consider the following initial state, a linear combination of one down 
spin states,  
  \begin{equation}
\vert\Psi(0)\rangle =  \alpha\vert10...0\rangle+\beta\vert010...0\rangle = 
(\alpha c^\dagger_1 + \beta c^\dagger_2) \prod_{q > 0} \vert0\rangle_q 
\vert0\rangle_{-q}.
 \end{equation} 

Time evolution of the state will generate all odd magnon sector states.
 From the explicit forms of the correlations functions given in Appendix 
\ref{Appendix-I} Eq.~\ref{correlation_fn_2} we can compute the bipartite 
measures of quantum correlations as a function of time. The nearest neighbour 
concurrence $C(i,i+1)$ and  the mutual information $I(i,i+1)$ are plotted as 
functions of site index $i$ and time $t$ for the XY model with a transverse 
field in Fig.~\ref{fig:6}. Here also we illustrate our results for the same 
three sets of Hamiltonian parameters $(J_x = 0.7, J_y = 0.3, h = 0.1)$; $(J_x = 
0.7, J_y = 0.3, h = 1.0)$ and $(J_x = 0.7, J_y = 0.3, h = 10.0)$. Unlike the 
Heisenberg model, the pairwise concurrence from the Bell state does not spread 
from first two sites for the cases $h = 0.1$ and $h = 1.0$ as shown in 
Fig.~\ref{fig:6}(a) and \ref{fig:6}(b). Since parity is conserved and the state 
contains all odd number of down spins, initially the pairwise concurrence is 
generated but after sometime it decays.  Similarly, the pairwise mutual 
information does not spread from first two sites for the cases $h = 0.1$ and $h 
= 1.0$ as shown in Fig.~\ref{fig:6}(d) and \ref{fig:6}(e) respectively. However, 
  the spins generate mutual information between them as a result of the 
dynamics.  But only for the case $h = 10.0$, the pairwise concurrence and the 
mutual information do spread with a finite speed from the first two sites and 
their values are zero outside the light cone as shown in Fig.~\ref{fig:6}(c) and 
\ref{fig:6}(f) respectively.\\

The transverse field XY model reduces to the Ising model when $J_y = 0$ and $h = 
0$. In this case the expectation values of the spin correlation functions take 
simpler and closed forms given in Appendix \ref{Appendix-II}  Eq.~ 
\ref{correlation_fn_3}.  The nearest neighbour concurrence $C(i,i+1)$ and the 
mutual information $I(i,i+1)$ are plotted as functions of site index $i$ and 
time $t$ for the Ising model in Fig.~\ref{fig:7}(a) and \ref{fig:7}(b) 
respectively.  Since the Ising dynamics entangles only two neighbour spins,  the 
local bipartite correlations and the entanglement at the first two sites 
trivially do not spread, and a speed of the correlation propagation cannot be 
defined. However, pairwise correlations between the spins are generated due to 
local two qubit entanglement operations.

The time dependence of tripartite mutual information (TMI) for the transverse field 
XY model and the Ising model has been plotted in  Fig.~\ref{fig:8}. In 
Fig.~\ref{fig:8}(a), the quantity $I(1:2:3)$ is plotted as a function of time 
$t$ and magnetic field $h$ for the XY model with a transverse field for the 
parameters $J_x = 0.7, J_y = 0.3$. The quantity $I(1:2:3)$ is mostly negative in 
the range $\vert h\vert< 1$ and non negative for large $h$ region.  It can also 
be seen in Fig.~\ref{fig:8}(b) that $I(1:2:3)$ becomes negative for the Ising 
model with parameters $J_x = 1.0, J_y = 0.0$ for the magnetic field in the range 
$\vert h\vert< 1$. For the Ising model without a transverse field, the value of 
the quantity $I(1:2:3)$ is strictly bounded between $0$ and $-1$ implying a 
perfect scrambling of information. For a higher value of the transverse field 
$\vert h\vert >> 1$, the value is again non negative. So, scrambling mostly 
takes place in the range  $\vert h\vert< 1$ for both these two non 
spin-conserving models. \textcolor{black} {For comprehensiveness of our analysis, 
the results for the Bell state with zero and two magnon subspace, i.e., $(\vert 
00...0\rangle + \vert 110...0\rangle)/\sqrt{2}$ for the  transverse field XY 
model and the Ising model have also been shown  in Fig.~\ref{fig:8}(b) and (c) 
respectively. Here all even magnon sectors will be generated unlike the  one 
magnon initial state.  The result is qualitatively similar to that of the one 
magnon initial state for the range $\vert h\vert< 1$. However, the quantity 
$I(1:2:3)$ is not strictly non negative in the limit $\vert h\vert >> 1$ unlike 
the one magnon initial state. This can be well understood, as in the said limit, 
the dynamics is not restricted to the a particular subspace, rather still mixes 
zero and two magnon sectors. } But in both the cases, the scrambling is much 
more prominent compared to that of the anisotropic Heisenberg model even with 
the initial state $(\vert00...0+ \vert110...0\rangle)/\sqrt{2}$ (Cf. 
Fig.~\ref{fig:3}). Therefore, it can be concluded that the confinement and decay 
of bipartite quantum correlations in local spins is associated with a high 
negative value of TMI as seen in the cases of the Ising model and the XY model 
with a small magnetic field.

\begin{sidewaystable}
\sidewaystablefn%
\begin{center}
\begin{minipage}{\textheight}
\begin{tabular*}{\textheight}{@{\extracolsep{\fill}}lcccccc@{\extracolsep{\fill}
}}
\hline
Spin-models & Spin conserving & Integrable & Sign of TMI & Finite speed of\\ 
{~~~~~~} & {~~~~~~} & {~~~~~~} & {~~~~~~} &  correlation propagation\\
\hline
Heisenberg model & yes & yes & non-negative & yes\\ 
\hline
Transverse field XY model ($\vert h\vert << J_x,J_y$) & no & yes & negative & no 
\\ 
\hline
Transverse field XY model ($h >> J_x,J_y$) & yes & yes & non-negative & yes \\  
\hline
Ising model & no & yes & negative & no\\ 
\hline
Kicked Harper model & yes & no & non-negative & yes\\ 
\hline
\end{tabular*}
\caption{The sign of the tripartite mutual information and the propagation of 
bipartite correlations through the spin chain for different model for one-magnon 
initial states. }
\label{tab1}
\end{minipage}
\end{center}
\end{sidewaystable}

 \section{Conclusions}
\label{section_6}

We have studied the dynamics of  spin chains for various model Hamiltonians and 
found some interesting generic interrelationships between bipartite quantum 
correlations and information scrambling. We start with a simple initial state,  
an entangled pair at the first two qubits, and then let the system evolve 
unitarily. The bipartite quantum correlations have been calculated  to show how 
the information coded in the first two sites as a form of quantum correlation 
spreads out to other locations. We have shown that the spreading of quantum 
correlations takes place when the dynamics is restricted to a subspace of the 
full Hilbert space. In the cases of the Heisenberg model, the transverse field 
XY model with a very large transverse field and the kicked Harper model, where 
the number of down (up) spins is a conserved quantity, the entanglement spreads 
consistently to other parts of the system. On the other hand,  for the XY model 
and the Ising model with a small and moderate transverse field $(\vert h\vert< 
1)$, where the dynamics is spin non-conserving, the propagation or spreading do 
not take place from the first two entangled sites, although quantum correlations 
are generated in the system.

This behaviour can be understood from the perspective of 
quantum information scrambling. If the dynamics is restricted to a spin 
conserving subspace, scrambling does not take place. We have shown that the 
Heisenberg model, the kicked Harper model and the transverse field XY model with 
$\vert h\vert >> 1$ do not show any scrambling behaviour for one down-spin 
states as TMI is always non-negative. In these cases, the bipartite correlations 
in the system do not delocalize into multiparitite correlations as time evolves. 
 On the other hand, the XY model with a small transverse field $(\vert h\vert< 
1)$, and the Ising model show a scrambling behaviour even with one down (up) 
spin initial states. We have also seen that quantum integrability does not play 
any direct role in scrambling the quantum correlations. The kicked Harper model, 
a nonintegrable model, does not show scrambling behaviour, whereas the XY model 
with a transverse field, an integrable model shows scrambling behaviour. In 
these cases, bipartite correlations quickly delocalize into multipartite 
correlations as time evolves. However, for other initial states that mix 
different spin sectors, scrambling does take place in spin-conserving dynamics. 
For example, the Heisenberg dynamics and the kicked Harper dynamics with an 
initial state with zero and two magnon sectors do show scrambling; though much 
smaller compared to that of spin non-conserving models. Occurrence of scrambling 
in a many-body quantum system, therefore, depends fundamentally on two factors: 
whether the bipartite correlations delocalize with time owing to non 
spin-conserving dynamics and whether the dynamics mixes different spin sectors.
 \\

Our main results are summarized in Table~\ref{tab1}. The table gives a clear 
relationship between the sign of tripartite mutual information and the 
propagation of bipartite  quantum correlations with time.  The models where 
locally encoded bipartite quantum correlations propagate with a finite speed 
have a non-negative value of TMI. On the other hand,  the information is lost 
and can not propagate in the models with a negative value of TMI.

\backmatter

\bmhead{Acknowledgments}
We acknowledge the support of Department of Physics, Indian Institute of 
Technology, Kanpur, where part of the work was carried out. VS acknowledges the 
support of Science and Engineering Research Board, through MATRICS scheme.

\bmhead{Data Availability}
All data generated or analyzed during this study are available from the authors 
on reasonable request.

\bigskip


\begin{appendices}

\section{Pairwise correlation functions for the XY model with transverse 
field}\label{Appendix-I}

The expectation values of the correlation functions $\langle 
c^{\dagger}_jc_j\rangle$, $\langle c_jc^{\dagger}_{j+1}\rangle$, $\langle 
c_jc_{j+1} \rangle$ and $\langle 
c^{\dagger}_jc_jc^{\dagger}_{j+1}c_{j+1}\rangle$ as a function of time can be 
calculated analytically as shown as,

\begin{center}
\begin{eqnarray}
&&\langle c^{\dagger}_jc_j\rangle_t = \frac{1}{N} \sum_{q_1,q_2} e^{iq_1j-iq_2j} 
\langle(\chi^*_{q_1}  c^{\dagger}_{q_1} + \xi^*_{q_1} c_{-q_1}) (\chi_{q_2}  
c_{q_2} + \xi_{q_2} c_{-q_2}^\dagger)\rangle,\nonumber\\
&&\langle c_jc_{j+1}\rangle_t =  \frac{1}{N}\sum_{q_1,q_2} e^{iq_1j-iq_2(j+1)}  
\langle(\chi_{q_1}  c_{q_1} + \xi_{q_1} c_{-q_1}^\dagger) (\chi_{q_2}  c_{q_2} + 
\xi_{q_2} c_{-q_2}^\dagger)\rangle,\nonumber\\
&&\langle c_jc^{\dagger}_{j+1}\rangle_t =  \frac{1}{N}\sum_{q_1,q_2} 
e^{-iq_1j+iq_2(j+1)}\langle(\chi_{q_1}  c_{q_1} + \xi_{q_1} 
c_{-q_1}^\dagger)(\chi^*_{q_2}  c^{\dagger}_{q_2} + \xi^*_{q_2} 
c_{-q_2})\rangle,\nonumber\\
&&\langle c^{\dagger}_jc_jc^{\dagger}_{j+1}c_{j+1}\rangle_t =  
\frac{1}{N^2}\sum_{q_1,q_2,q_3,q_4}e^{iq_1j-iq_2j+iq_3(j+1)-iq_4(j+1)} 
\langle(\chi^*_{q_1}  c^{\dagger}_{q_1} + \xi^*_{q_1} c_{-q_1})\nonumber\\ && 
{~~~~~~~~~~~~~~~~~~~~~~~}(\chi_{q_2}  c_{q_2} + \xi_{q_2} c_{-q_2}^\dagger) 
(\chi^*_{q_3} c^{\dagger}_{q_3} +\xi^*_{q_3} c_{-q_3})(\chi_{q_4}  c_{q_4} + 
\xi_{q_4}c_{-q_4}^\dagger)\rangle.\nonumber\\
\label{correlation_fn_2}
\end{eqnarray}
\end{center}

\begin{figure*}[t]
     \begin{center}

        \subfigure[]{%
           \includegraphics[width=0.26\textwidth]{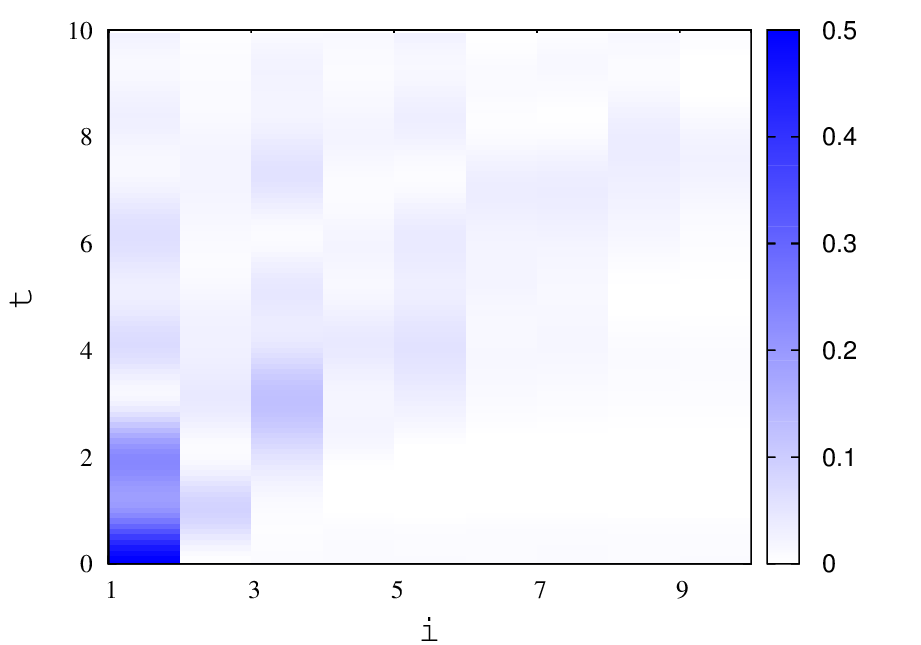}
       }%
       \subfigure[]{%
          \includegraphics[width=0.26\textwidth]{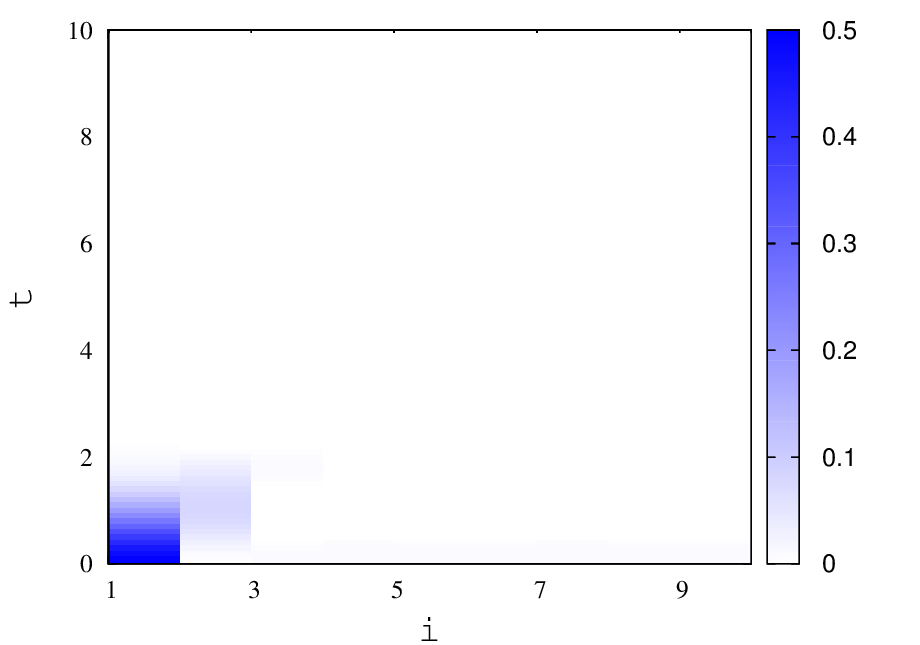}
       }%
       \subfigure[]{%
           \includegraphics[width=0.26\textwidth]{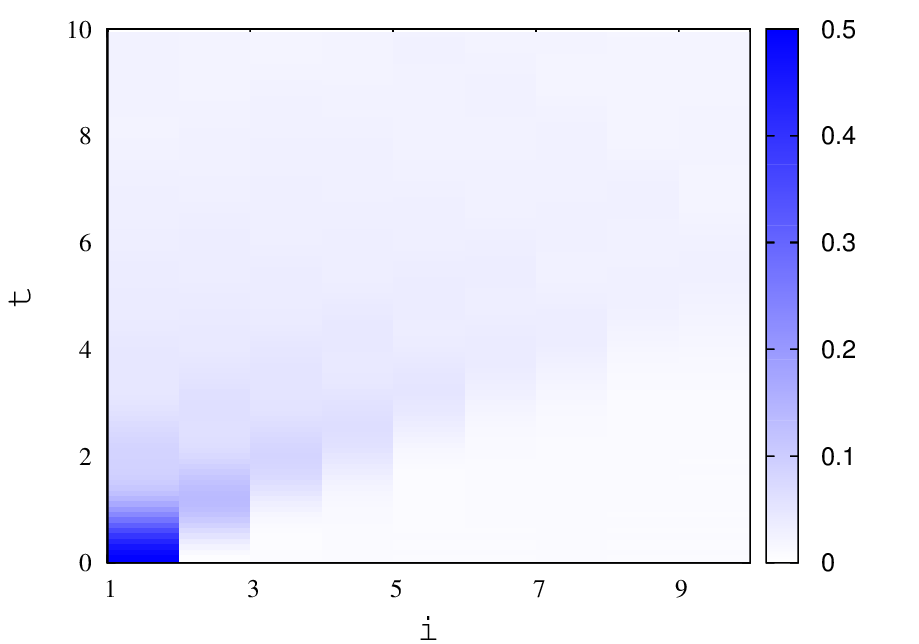}
       }\\%
    \subfigure[]{%
       \includegraphics[width=0.26\textwidth]{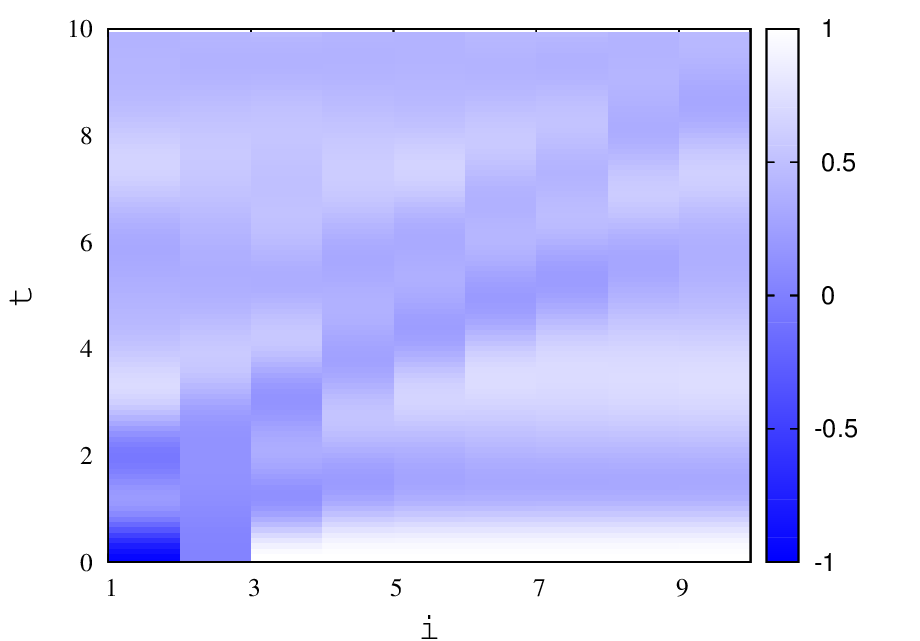}
    }%
    \subfigure[]{%
       \includegraphics[width=0.26\textwidth]{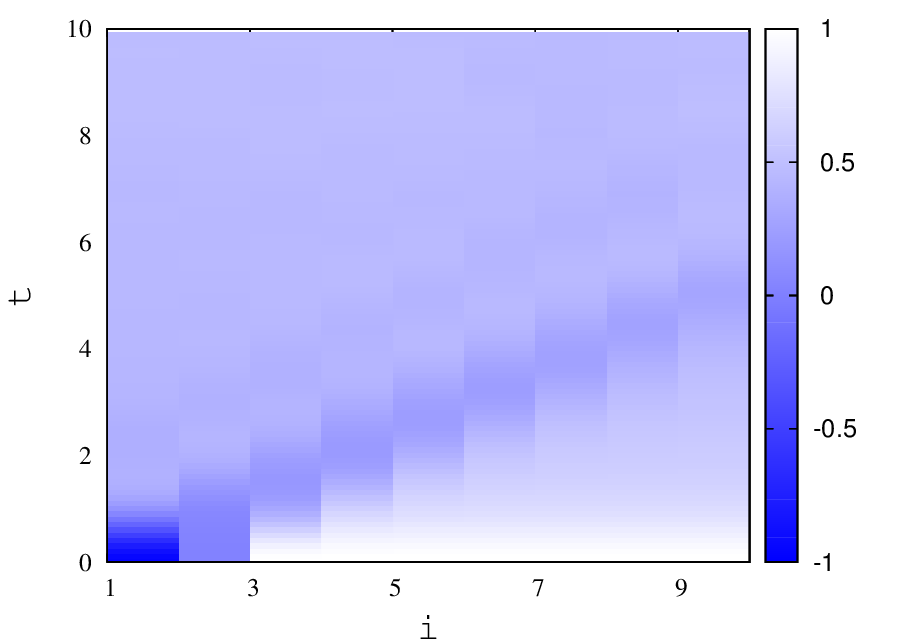}
    }%
    \subfigure[]{%
       \includegraphics[width=0.26\textwidth]{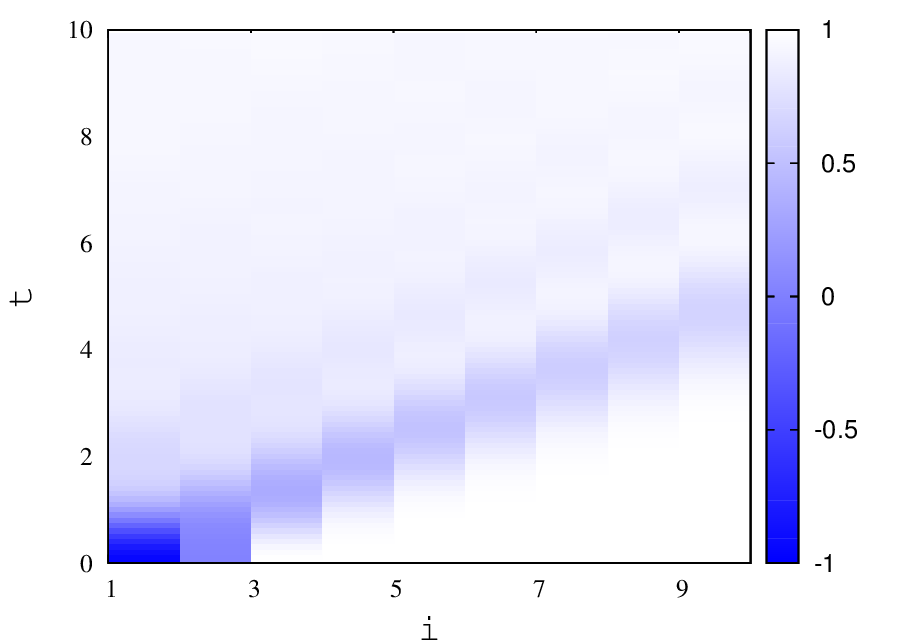}
        }\\%
       \subfigure[]{%
             \includegraphics[width=0.26\textwidth]{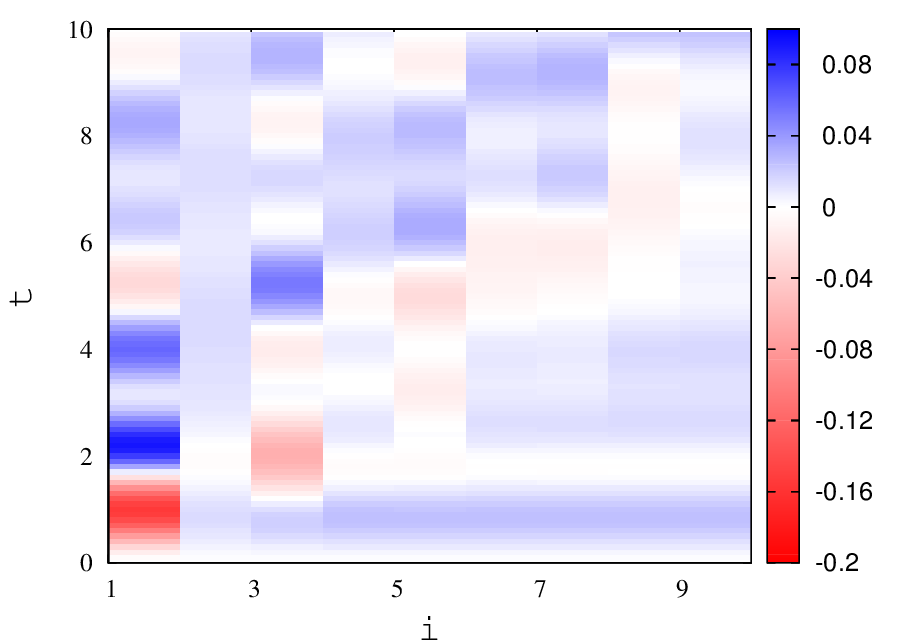}
        }%
        \subfigure[]{%
           \includegraphics[width=0.26\textwidth]{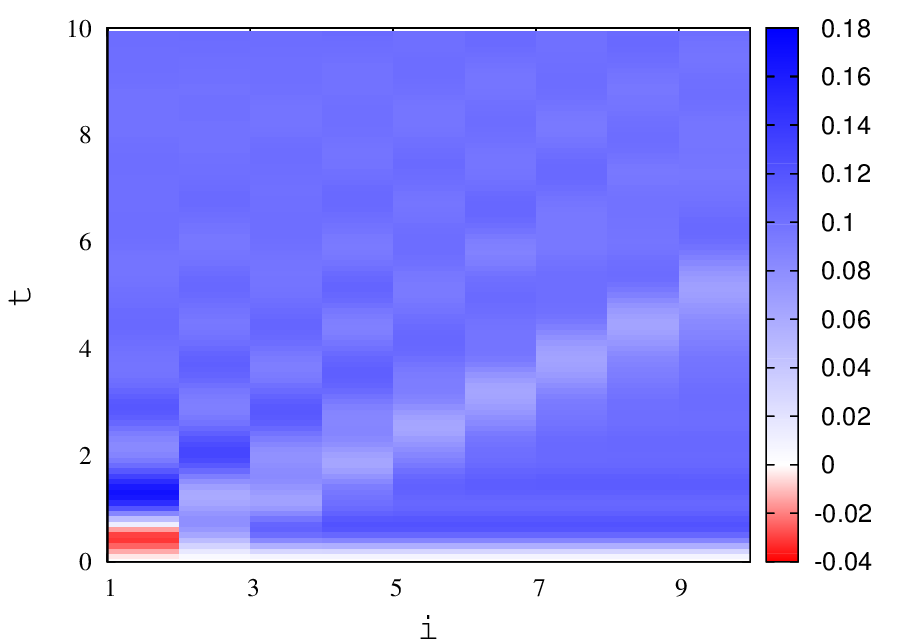}
        }%
       \subfigure[]{%
          \includegraphics[width=0.26\textwidth]{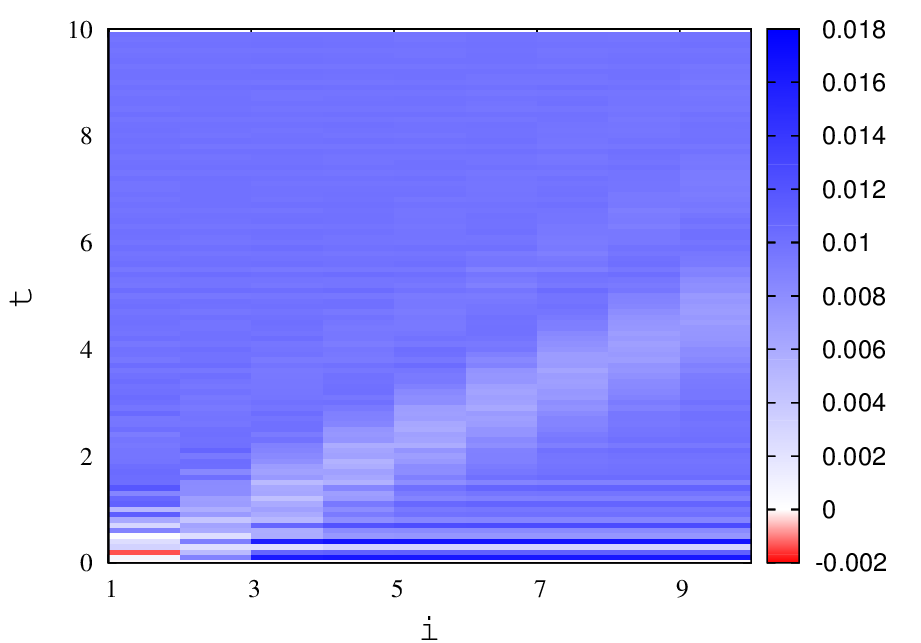}
       }\\
   \end{center}
\caption{{The correlation functions between the sites $i$ and $i+1$ are shown as 
functions of time $t$ for the transverse field XY model for various Hamiltonian 
parameters.  $\rm Re \langle \sigma^+_i \sigma^-_{i+1}\rangle_t$ is plotted as a 
function of time for parameters: (a) $J_x =0.7, J_y = 0.3$ and $h = 0.1$,  (b)  
$J_x =0.7, J_y = 0.3$ and $h = 1.0$,  (c)   $J_x =0.7, J_y = 0.3$ and $h = 
10.0$. $\rm Re \langle \sigma^+_i \sigma^+_{i+1}\rangle_t$ is plotted as a 
function of time for parameters: (d) $J_x =0.7, J_y = 0.3$ and $h = 0.1$,  (e)  
$J_x =0.7, J_y = 0.3$ and $h = 1.0$,  (f)   $J_x =0.7, J_y = 0.3$ and $h = 
10.0$.  $\langle \sigma^z_i \sigma^z_{i+1}\rangle_t$ for parameters: (g) $J_x 
=0.7, J_y = 0.3$ and $h = 0.1$,  (h)  $J_x =0.7, J_y = 0.3$ and $h = 1.0$,  (i)  
 $J_x =0.7, J_y = 0.3$ and $h = 10.0$. The results are shown from analytical 
calculations for the initial state:  $\vert\Psi(0)\rangle = 
(\vert10...0\rangle+\vert010...0\rangle)/\sqrt{2}$. }
     }%
  \label{fig:9}
\end{figure*}

We see from the above set of equations that the time evolution mixes the 
different operators. Any product of fermion operators in the real space involves 
$N$ sums over momenta values in real space. The expectation values of products 
of fermionic operators can be straightforwardly evaluated using Wick's theorem 
\cite{peskin, lindgren}. However, there will be $N$ sums over the momentum 
variables. For a large value of  $N$, the above sums are evaluated by converting 
them to integrals from $0$ to $\pi$.  The pairwise correlation functions between 
the nearest neighbours are plotted as functions of site index $i$ and time $t$ 
in Fig.~\ref{fig:9}. We have taken three sets of Hamiltonian parameters $(J_x = 
0.7, J_y = 0.3, h = 0.1)$; $(J_x = 0.7, J_y = 0.3, h = 1.0)$ and $(J_x = 0.7, 
J_y = 0.3, h = 10.0)$ to illustrate the results. The total number of down (up) 
spins in the system is not conserved as $J_x \neq J_y$. However, in the limit $h 
\rightarrow \infty$, the dynamics is confined to a subspace of the total Hilbert 
space, as the Hamiltonian in this case commutes with the total number of down 
(up) spins. The initial state being $\vert\Psi\rangle = 
\frac{1}{\sqrt{2}}(100...0 + 010...0)$, the initial values of the off-diagonal 
correlation function $\langle \sigma^+_i \sigma^-_{i+1}\rangle$ is $0.5$  the 
diagonal correlation function $\langle \sigma^z_i \sigma^z_{i+1}\rangle$ is $-1$ 
for the first pair and zero for all other pairs.

As time evolves, the correlation function $\langle \sigma^+_i 
\sigma^-_{i+1}\rangle$ becomes non zero for farther sites for later times  
implying a finite speed of propagation of correlations) for the case $h = 0.1$ 
as shown in Fig.~\ref{fig:9}(a). The value of the function $\langle \sigma^+_i 
\sigma^-_{i+1} \rangle$ is non zero  within the `light cone' but zero outside. 
For the case $h = 1.0$, the correlation function $\langle \sigma^+_i 
\sigma^-_{i+1} \rangle$ decays very quickly and becomes zero beyond the third 
site as shown in Fig.~\ref{fig:9}(b). For the case $h = 10.0$, the correlations  
propagate consistently and continuously to further sites with a finite speed as 
shown in Fig.~\ref{fig:9}(c). The diagonal correlation function $\langle 
\sigma^z_i \sigma^z_{i+1} \rangle$ becomes non zero for farther sites quickly 
and propagation does not take place with a finite speed for the cases $h = 0.1$ 
and $h = 1.0$ as shown in Fig.~\ref{fig:9}(d) and Fig.~\ref{fig:9}(e) 
respectively. For the case $h = 10.0$, the value of the function $\langle 
\sigma^z_i \sigma^z_{i+1} \rangle$ spreads with finite speed and its value  is 
zero outside the light cone as shown Fig.~\ref{fig:9}(f). The correlation 
function $\langle \sigma^+_i \sigma^+_{i+1} \rangle$ is plotted as a function of 
time and site index for the same set of Hamiltonian parameters in 
Fig.~\ref{fig:9}(g), (h) and (i) respectively. The expectation value of the 
correlation function $\langle \sigma^+_i \sigma^+_{i+1} \rangle$ is initially 
zero and becomes non zero as the number of down spins increases in the system. 
The values  of  $\langle \sigma^+_i \sigma^+_{i+1} \rangle$  for the case $h = 
10.0$ in Fig.~\ref{fig:9}(i) is  much smaller compared to the cases $h = 0.1$ 
and $h = 1.0$ as seen in Fig.~\ref{fig:9}(g) and Fig.~\ref{fig:9}(h) 
respectively. This indicates that for a high value of $h$, less number of down 
spins are generated in the system as time evolves and the dynamics remains 
confined mainly in the one spin sector.

\setcounter{equation}{0}

\section{Pairwise correlation functions for the Ising model}\label{Appendix-II}

For the Ising model the time evolved expectation value of any operator $O$ is 
given as
\begin{eqnarray}
\langle O \rangle_t = \langle e^{iJt\sum_i \sigma^x_{i} \sigma^x_{i+1}} O 
e^{-iJt\sum_i \sigma^x_{i} \sigma^x_{i+1}} \rangle.
 \end{eqnarray}
The  forms of $\langle \sigma^z_j \rangle_t, \langle \sigma^\pm_j\rangle_t, 
\langle \sigma^z_j \sigma^z_{j+1}\rangle_t $ and $\langle \sigma^+_j 
\sigma^\mp_{j+1}\rangle_t$ can be given by the following,

 \begin{center}
 \begin{eqnarray}
&&\langle \sigma^z_j\rangle_t = (C^4 +S^4 - 2C^2 S^2)\langle \sigma^z_j\rangle  
+ (2CS^3- 2C^3S) (\langle \sigma^y_j \sigma^x_{j+1}\rangle + \langle 
\sigma^x_{j-1} \sigma^y_{j}\rangle) 
\nonumber\\&&{~~~~~~~~~~~~~~~~~~~~~~~~~~~~~~~~~~~~~~~~~~~~~~}  -4C^2S^2 \langle 
\sigma^x_{j-1} \sigma^z_{j} \sigma^x_{j+1}\rangle,\nonumber\\
&&\langle \sigma^\pm_j\rangle_t = 0,\nonumber\\
&&\langle \sigma^z_j \sigma^z_{j+1}\rangle_t= (C^6 +S^6 - C^4S^2 -C^2S^4)\langle 
\sigma^z_j \sigma^z_{j+1}\rangle + (CS^5+C^3S^3-2C^5S) (\langle \sigma^z_j 
\sigma^y_{j+1}\sigma^x_{j+2}\rangle  \nonumber\\&& + \langle \sigma^x_{j-1} 
\sigma^y_{j}\sigma^z_{j+1}\rangle)+ (4C^2S^4 +4C^4S^2) \langle \sigma^x_{j-1} 
\sigma^y_{j} \sigma^y_{j+1} \sigma^x_{j+2} \rangle,\nonumber\\
&&\langle \sigma^+_j \sigma^\mp_{j+1}\rangle_t= (C^6 +S^6)\langle \sigma^+_j 
\sigma^\mp_{j+1}\rangle + (C^4S^2 +C^2S^4) (\langle \sigma^+_j 
\sigma^\pm_{j+1}\rangle + \langle \sigma^-_j \sigma^\pm_{j+1}\rangle + \langle 
\sigma^-_j \sigma^\mp_{j+1}\rangle) \nonumber\\&&+ (iC^5S - iC^3 S^3) 
(\frac{1}{2}\langle \sigma^+_j (1\mp\sigma^z_{j+1}) \sigma^x_{j+2} \rangle + 
\frac{1}{2}\langle \sigma^x_{j-1} (1+\sigma^z_{j}) \sigma^\mp_{j+1} \rangle + 
\frac{1}{4}(1+\sigma^z_{j})(1\mp\sigma^z_{j+1}))\nonumber\\&&+ (iCS^5 - iC^5 S) 
(\frac{1}{2}\langle \sigma^+_j (1\pm\sigma^z_{j+1}) \sigma^x_{j+2} \rangle + 
\frac{1}{2}\langle \sigma^x_{j-1} (1-\sigma^z_{j+1}) \sigma^\mp_{j+1} \rangle + 
\frac{1}{4}(1-\sigma^z_{j})(1\pm\sigma^z_{j+1}))\nonumber\\
&& +(C^2S^4+C^4S^2)(\frac{1}{4} \langle\sigma^x_{j-1} 
(1+\sigma^z_{j})(1\pm\sigma^z_{j+1})\sigma^x_{j+2}\rangle  -\frac{1}{4} 
\langle\sigma^x_{j-1} (1-\sigma^z_{j})(1\pm\sigma^z_{j+1})\sigma^x_{j+2}\rangle 
\nonumber\\&&
-\frac{1}{4} \langle\sigma^x_{j-1} 
(1+\sigma^z_{j})(1\mp\sigma^z_{j+1})\sigma^x_{j+2}\rangle 
+\frac{1}{4} \langle\sigma^x_{j-1} 
(1-\sigma^z_{j})(1\mp\sigma^z_{j+1})\sigma^x_{j+2}\rangle  \nonumber\\ && + 
\frac{1}{2} \langle\sigma^x_{j-1}\sigma^-_{j} (1\mp\sigma^z_{j+1})\rangle
 + \frac{1}{2} \langle\sigma^x_{j-1}\sigma^-_{j} (1\pm\sigma^z_{j+1})\rangle  - 
\frac{1}{2} \langle\sigma^x_{j-1}\sigma^+_{j} (1\pm\sigma^z_{j+1})\rangle   
\nonumber\\ &&- \frac{1}{2} \langle\sigma^x_{j-1}\sigma^+_{j} 
(1\mp\sigma^z_{j+1})\rangle)  +\frac{1}{2} \langle 
(1+\sigma^z_{j})\sigma^\pm_{j+1}\sigma^x_{j+2} \rangle + \frac{1}{2} \langle 
(1+\sigma^z_{j})\sigma^\pm_{j+1}\sigma^x_{j+2} \rangle \nonumber\\&&- 
\frac{1}{2} \langle (1-\sigma^z_{j})\sigma^\pm_{j+1}\sigma^x_{j+2} \rangle  
-\frac{1}{2} \langle (1+\sigma^z_{j})\sigma^\pm_{j+1}\sigma^x_{j+2} \rangle), 
\nonumber\\
 \label{correlation_fn_3}
 \end{eqnarray}
 \end{center}

with, $C \equiv \cos(Jt)$ and $S \equiv \sin(Jt)$.




\end{appendices}

\end{document}